\documentclass[epj]{svjour}
\usepackage{graphics}
\usepackage[T1]{fontenc}
\usepackage[latin1]{inputenc}
\usepackage{amsmath}
\usepackage{graphicx}
\usepackage{amssymb}

\usepackage{graphicx}
\usepackage{epsfig}
\usepackage{multirow}
\usepackage{dcolumn}
\usepackage{bm}
\usepackage{subfigure}

\def\<{\langle}
\def\>{\rangle}

\begin{document}

\title{The spectrum of interacting metallic carbon nanotubes: Exchange effects and universality}
\author{Leonhard Mayrhofer \and Milena Grifoni% etc
}   

\institute{Theoretische Physik, Universit\"{a}t Regensburg, 93040 Regensburg, Germany}
\date{Received: date / Revised version: date}
\abstract{
The low energy spectrum of finite size metallic single-walled carbon nanotubes (SWNTs) is determined. Starting
from a tight binding model for the $p_{z}$ electrons, we derive the
low energy Hamiltonian containing \textit{all relevant scattering processes}
resulting from the Coulomb interaction, including the short
ranged contributions becoming relevant for small diameter tubes. In
combination with the substructure of the underlying honeycomb lattice
the short ranged processes lead to various exchange effects. Using
bosonization the spectrum is determined. We find that the ground state
is formed by a spin $1$ triplet, if $4n+2$ electrons occupy the
SWNT and the branch mismatch is smaller than the exchange splitting.
Additionally, we calculate the excitation spectra for the different
charge states and find the lifting of spin-charge separation as well
as the formation of a quasi-continuum at higher excitation energies.
 % end of PACS codes
} %end of abstract
\authorrunning{L. Mayrhofer and M. Grifoni}
\titlerunning{The spectrum of interacting metallic carbon nanotubes}
\maketitle

\section{Introduction}

Single walled carbon nanotubes (SWNTs) have remarkable mechanical
and electronic properties. They represent, at low enough energies,
an almost ideal realization of an one-dimensional (1D) electronic
system with an additional orbital degree of freedom. Due to this 1D
character the proper inclusion of the Coulomb interaction between
the electrons in a SWNT is mandatory. For metallic SWNTs of infinite
length the theoretical works \cite{Egger1997,Odintsov1999} showed that correlations between the electrons
can be described within the Luttinger liquid picture. The accompanying
occurrence of power-laws for various transport properties could indeed
be observed experimentally \cite{Bockrath1999,Postma2001}. The effects of the forward scattering part of the
electron-electron interactions in
finite-size SWNTs were treated by Kane et al. in \cite{Kane1997} within the bosonization framework. There
the discrete energy spectrum of the collective spin and charge excitations
was derived. The bosonization method has recently been used also to
determine the transport properties of finite size metallic SWNT quantum
dots \cite{Mayrhofer2006}.

So far the effect of non-forward scattering parts of the Coulomb interaction
has only been discussed for SWNTs of infinite length by renormalization
group techniques \cite{Egger1997,Odintsov1999}. In \cite{Egger1997}
deviations from conventional Luttinger Liquid behaviour have been
found only for very small temperatures $T\lesssim0.1\textrm{ mK}$
provided that the interaction is long ranged. The work of Odintsov
et al. \cite{Odintsov1999} additionally took into account the situation
at half filling where the formation of a Mott insulating state was
predicted. In the works treating electron-electron interactions in
finite size SWNTs within the bosonization formalism, the effect of
non-forward scattering parts of the Coulomb interaction has been neglected.
This approximation, which we will call {}``standard'' theory in
the following, is valid if moderate to large diameter tubes ($ \gtrsim 1.5 \textrm{ nm}$)
are considered as in \cite{Kane1997,Mayrhofer2006}, or if finite
size effects can be neglected since the relevant energies exceed the
level spacing of the SWNT as in the experiments \cite{Bockrath1999,Postma2001}.
Recent experiments \cite{Liang2002,Sapmaz2005,Moriyama2005} however
have found exchange effects in the ground state spectra of small diameter
tubes which can not be explained using the {}``standard'' bosonization
theory for interacting SWNTs. Oreg et al. \cite{Oreg2000} have presented
a mean-field Hamiltonian for the low energy spectrum of SWNTs including
an exchange term favouring the spin alignment of electrons in different
bands. The values for the exchange energies observed in the experiments
agree well with the mean-field predictions. However, the question of a singlet-triplet ground state is beyond the mean field
approach.
Moreover, in contrast to the bosonization procedure it can not predict
the strong energy renormalization of the charged collective electron
excitations.

In this article we go beyond the mean-field approach. We derive a low-energy
Hamiltonian for finite size metallic SWNTs, which includes \textit{all}
relevant short-ranged interaction processes. This allows us to identify
the microscopic mechanisms that lead to the various exchange effects.
Using bosonization we determine the spectrum and eigenstates of the
SWNT Hamiltonian essentially exactly away from half-filling. 
An interesting situation arises near half-filling since there additional
processes become relevant which can not be considered as small compared
to the dominating forward scattering terms. Unfortunately we have
not found a reliable way of diagonalizing the Hamiltonian in that
situation so far.

Concerning the ground state properties, we find
under the condition of degenerate or almost degenerate bands, a spin $1$ triplet as ground
state if $4n+2$ electrons occupy the nanotube. This is insofar remarkable
as a fundamental theorem worked out by Lieb and Mattis \cite{Lieb-Mattis1962}
states for any single-band Hubbard model in 1D with nearest-neighbour
hopping that the ground state can only have spin $0$ or $1/2.$ However
at the end of their article they explicitly pose the question whether
ground states with higher spin could be realized in 1D systems with
orbital degeneracy, which in the case of SWNTs is present due to the
substructure of the underlying honeycomb lattice. Our findings answer
this question with yes, hence proofing that the theorem by Lieb and
Mattis can not be generalized to multi-band systems. Moreover it is
interesting to notice that all of the processes favouring higher spin
states in SWNTs involve non-forward scattering with respect to the
orbital degree of freedom. On the experimental side an exchange splitting
in the low energy spectrum of the $4n+2$ charge state has indeed
been observed \cite{Liang2002,Sapmaz2005,Moriyama2005}. However,
all the experiments demonstrating exchange splitting were carried
out for SWNTs with a large band mismatch such that the ground states
are supposed to be spin $0$ singlets. Especially Moriyama et al.
have proven that this is the case in their experiments \cite{Moriyama2005}
by carrying out magnetic field measurements. Thus the threefold degenerate
spin $1$ ground state has not been observed yet, since its occurrence
requires a band mismatch that is small compared to the exchange energy.
Additionally to the ground state properties of metallic SWNTs we have
also determined the excitation spectra. We find that the huge degeneracies
as obtained by only retaining the forward scattering processes are
partly lifted and the spectrum becomes more and more continuous when
going to higher energies. Finally this leads to a lifting of the spin
charge separation predicted by the {}``standard'' theory. 

The outline of this article is the following. We start in Section
\ref{sec:The-noninteracting-System} by briefly reviewing the low
energy physics of noninteracting electrons in finite size metallic
SWNTs. Including the Coulomb interaction we derive the effectively
one-dimensional Hamiltonian for the low energy regime in Section \ref{sec:The-interaction-Hamiltonian}.
The subsequent examination of the effective 1D interaction potential
in Section \ref{sub:The-relevant-scattering} allows us to sort out
the irrelevant interaction processes. The remaining processes are
either of density-density or non-density-density form. The former
ones we diagonalize together with the kinetic part of the Hamiltonian
by bosonization in \ref{sub:Diagonalizing}.
Using the obtained eigenstates as basis we calculate the corresponding
matrix elements for the non-density-density part of the interaction
with the help of the bosonization identity of the electron operators,
Section \ref{sub:The-matrix-elements}. In Section \ref{sec:The-SWNT-spectrumresults}
we calculate the ground state and excitation spectra by diagonalizing
the Hamiltonian including the non-density-density processes in a truncated
basis and discuss the results.

For the hurried reader we propose to skip the more technical sections
and, after reading Section \ref{sec:Low-energy-Hamiltonian}, to go
directly to Sections \ref{sub:The-low-energy} and \ref{sub:Excitation-spectra-away}
where the results of this work are presented for the low energy and
low to intermediate energy regions, respectively.

\section{Low energy Hamiltonian of metallic finite size SWNTs\label{sec:Low-energy-Hamiltonian}}

As shown in \cite{Odintsov1999_2}, correlation effects in metallic
SWNTs are universal at low energies, i. e. they do not depend on the
chirality of the considered tube. Therefore we can, without loss of
generality, focus on armchair nanotubes from now on. 

In this section we will give a short summary of the electronic structure
of noninteracting finite size armchair nanotubes in the low energy
regime following our earlier work \cite{Mayrhofer2006}. On this basis
we are going to include the Coulomb interaction between the electrons,
leading to an effective 1D Hamiltonian. The subsequent examination
of the effective 1D interaction potential will determine all the relevant
scattering processes, which are either of density-density or non-density-density
form.

\subsection{The noninteracting System\label{sec:The-noninteracting-System}}

\begin{figure}
\begin{center}\includegraphics{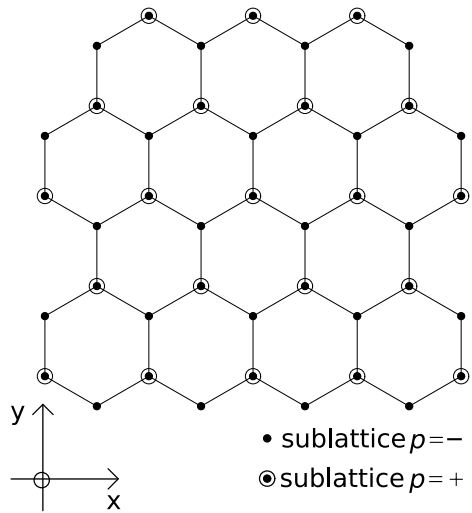}\end{center}

\caption{\label{cap:The-graphene-lattice}The graphene lattice with its sublattice
structure.}
\end{figure}

Before considering the effect of the electron-electron interactions,
let us recall the most important facts about noninteracting electrons
in finite size armchair SWNTs. Since SWNTs can be considered as graphene
sheets rolled up to cylinders, the bandstructure of SWNTs is easily
derived from the one of the $p_{z}$ electrons in the graphene honeycomb
lattice, see e.g. \cite{DiVizenco1984}. Two carbon atoms $p=\pm$
occupy the unit cell of graphene, cf. Fig. \ref{cap:The-graphene-lattice},
leading to a valence and a conduction band touching at the two Fermi
points $F=\pm K_{0}\hat{e}_{x}$. %
Quantization around the circumference of a SWNT restricts the set
of allowed wave vectors, leading to the formation of subbands. For
metallic SWNTs, only the gapless subbands with linear dispersion,
touching at the Fermi points, are relevant at low enough energies.
Imposing open boundary conditions along the tube length $L$, the
eigenfunctions of the noninteracting Hamiltonian $H_{0}$ are standing
waves $\varphi_{r\kappa}(\vec{r})$ where the occurrence of the branch
or pseudo spin index $r=\pm$ is a consequence of the double occupancy
of the graphene unit cell. Furthermore $\kappa$ measures the wave
number relative to the Fermi wave number $K_{0}$ and is subject to
the quantization condition \begin{equation}
\kappa=\frac{\pi}{L}(n_{\kappa}+\Delta),\, n_{\kappa}\in\mathbb{Z},\,\left|\Delta\right|\le1/2.\label{eq:QCond1}\end{equation}
The parameter $\Delta$ has to be introduced if there is no integer
$n$ with $K_{0}=\pi n/L,$ where $L$ is the tube length, and is
responsible for a possible energy mismatch $\varepsilon_{\Delta}$
between the $r=+$ and $r=-$ electrons. In general $\Delta$ depends
also on the type of the considered SWNT \cite{Jiang2002}. Explicitly,
$\varphi_{r\kappa}(\vec{r})$ can be decomposed into contributions
from the two sublattices $p=\pm,$\begin{equation}
\varphi_{r\kappa}(\vec{r})=\frac{1}{\sqrt{2}}\sum_{p=\pm}f_{pr}\left(e^{i\kappa x}\varphi_{pK_{0}}(\vec{r})-e^{-i\kappa x}\varphi_{p-K_{0}}(\vec{r})\right).\label{eq:phi_r_phi_p}\end{equation}
The coefficients $f_{pr}$ are given by \begin{equation}
f_{pr}=\left\{ \begin{array}{cc}
1/\sqrt{2}, & \, p=+\\
-r/\sqrt{2}, & \, p=-\end{array}\right.,\label{eq:fpr}\end{equation}
and the functions $\varphi_{pF}$ describe fast oscillating Bloch
waves on sublattice $p$ at the Fermipoint $F$, \begin{equation}
\varphi_{pF}(\vec{r})=\frac{1}{\sqrt{N_{L}}}\sum_{\vec{R}}e^{iFR_{x}}\chi(\vec{r}-\vec{R}-\vec{\tau}_{p}),\label{eq:phi_pF_chi}\end{equation}
 where $N_{L}$ is the total number of lattice sites and $\chi(\vec{r}-\vec{R}-\vec{\tau}_{p})$
is the $p_{z}$ orbital localized on site $\vec{R}$ of sublattice
$p$, see Fig. \ref{cap:The-graphene-lattice}.

\begin{figure}
\begin{center}\includegraphics[%
  width=0.45\columnwidth]{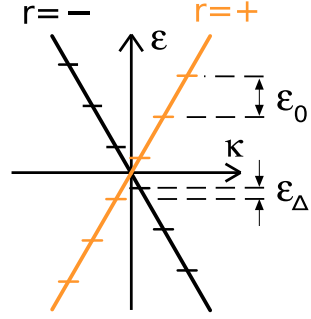}\end{center}

\caption{\label{cap:dispersion}The energy spectrum of a noninteracting metallic
SWNT with the two branches $r=\pm$. The level spacing is denoted
$\varepsilon_{0}$ and $\varepsilon_{\Delta}$ is the energy mismatch
between $r=+$ and $r=-$.}
\end{figure}

In Fig. \ref{cap:dispersion} we show the linear dispersion relation
for the standing waves $\varphi_{r\kappa}$. The slopes of the two
branches are given by $r\hbar v_{F},$ with the Fermi velocity $v_{F}\approx8.1\cdot10^{5}m/s$.
Including the spin degree of freedom, the Hamiltonian of the noninteracting
system $H_{0}$ therefore reads\begin{equation}
H_{0}=\hbar v_{F}\sum_{r\sigma}r\sum_{\kappa}\kappa c_{r\sigma\kappa}^{\dagger}c_{r\sigma\kappa},\label{eq:H_0_fermion}\end{equation}
where $c_{r\sigma\kappa}$ annihilates an electron in the state$\left|\varphi_{r\kappa}\right\rangle \left|\sigma\right\rangle $.
Thus the level spacing of the noninteracting system is given by \begin{equation}
\varepsilon_{0}=\hbar v_{F}\frac{\pi}{L}.\label{eq:levelspacing_nonint}\end{equation}

In the next section we are going to express the interaction part of
the Hamiltonian in terms of the 3D electron operators, which expressed
in terms of the wave functions $\varphi_{r\kappa}(\vec{r})$ read
\[
\Psi(\vec{r})=\sum_{\sigma}\sum_{r\kappa}\varphi_{r\kappa}(\vec{r})c_{r\sigma\kappa}=:\sum_{\sigma}\Psi_{\sigma}(\vec{r}).\]
By defining the slowly varying 1D electron operators, \[
\psi_{rF\sigma}(x):=\frac{1}{\sqrt{2L}}\sum_{\kappa}e^{i\mathrm{sgn}(F)\kappa x}c_{r\sigma\kappa},\]
 we obtain with (\ref{eq:phi_r_phi_p}),\begin{equation}
\Psi_{\sigma}(\vec{r})=\sqrt{L}\sum_{rF}\mathrm{sgn}(F)\psi_{rF\sigma}(x)\sum_{p}f_{pr}\varphi_{pF}(\vec{r}).\label{eq:3D_1D}\end{equation}

\subsection{\label{sec:The-interaction-Hamiltonian}The interaction Hamiltonian}

In this section we examine the interaction part of the Hamiltonian.
After introducing an effective 1D interaction potential, we discuss
which of the scattering processes are of importance. We start with
the general expression for the Coulomb interaction,\[
V=\frac{1}{2}\sum_{\sigma\sigma'}\int d^{3}r\int d^{3}r'\Psi_{\sigma}^{\dagger}(\vec{r})\Psi_{\sigma'}^{\dagger}(\vec{r}')U(\vec{r}-\vec{r}')\Psi_{\sigma'}(\vec{r}')\Psi_{\sigma}(\vec{r}),\]
 where $U(\vec{r}-\vec{r}')$ is the Coulomb potential. For the actual
calculations we model $U(\vec{r}-\vec{r}\,')$ by the so called Ohno
potential which interpolates between $U_{0},$ the interaction energy
between two $p_{z}$ electrons in the same orbital for $\vec{r}=\vec{r}\,'$
and $\frac{e^{2}}{4\pi\epsilon_{0}\epsilon\left|\vec{r}-\vec{r}\,'\right|}$
for large values of $\left|\vec{r}-\vec{r}\,'\right|.$ Measuring
distances in units of $\textrm{\AA{}}$ and energy in $\mathrm{eV}$,
it is given by \cite{Barford} \begin{equation}
U(\vec{r}-\vec{r}\,')=U_{0}/\sqrt{1+\left(U_{0}\epsilon\left|\vec{r}-\vec{r}\,'\right|/14.397\right)^{2}}\,\mathrm{eV}.\label{eq:Ohno}\end{equation}
 A reasonable choice is $U_{0}=15$ eV \cite{Fulde1995}. The dielectric
constant is given by $\epsilon\approx1.4-2.4$ \cite{Egger1997}.
Reexpressing the 3D electron operators $\Psi_{\sigma}(\vec{r})$ in
terms of the 1D operators $\psi_{rF\sigma}(x)$, cf. equation (\ref{eq:3D_1D}),
and integrating over the coordinates perpendicular to the tube axis,
we obtain, \begin{multline}
V=\frac{1}{2}\sum_{\sigma\sigma'}\sum_{\{[r],[F]\}}\mathrm{sgn}(F_{1}F_{2}F_{3}F_{4})\int dx\int dx'U_{[r][F]}(x,x')\\
\times\psi_{r_{1}F_{1}\sigma}^{\dagger}(x)\psi_{r_{2}F_{2}\sigma'}^{\dagger}(x')\psi_{r_{3}F_{3}\sigma'}(x')\psi_{r_{4}F_{4}\sigma}(x),\label{eq:V1D_rs}\end{multline}
where $\sum_{\{[r],[F]\}}$ denotes the sum over all quadruples $[r]=(r_{1},r_{2},r_{3},r_{4})$
and $[F]=(F_{1},F_{2},F_{3},F_{4})$. Under the assumption, justified
by the localized character of the $p_{z}$ orbitals, that the sublattice
wave functions $\varphi_{pF}(\vec{r})$ and $\varphi_{-pF}(\vec{r})$
do not overlap, i.e., $\varphi_{pF}(\vec{r})\varphi_{-pF}(\vec{r})\equiv0$,
the effective 1D Coulomb potential $U_{[r][F]}(x,x')$ is given by,
\begin{multline}
U_{[r][F]}(x,x')=L^{2}\int d^{2}r_{\perp}\int d^{2}r'_{\perp}\sum_{p,p'}f_{pr_{1}}f_{p'r_{2}}f_{p'r_{3}}f_{pr_{4}}\\
\times\varphi_{pF_{1}}^{*}(\vec{r})\varphi_{p'F_{2}}^{*}(\vec{r}')\varphi_{p'F_{3}}(\vec{r}')\varphi_{pF_{4}}(\vec{r})U(\vec{r}-\vec{r}').\label{eq:U_rF_origin}\end{multline}
 Using relation (\ref{eq:fpr}) for the coefficients $f_{pr}$ and
performing the sum over $p,p'$, we can separate $U_{[r][F]}$ into
a part describing the interaction between electrons living on the
same (intra) and on different (inter) sublattices, \begin{multline}
U_{[r][F]}(x,x')=\frac{1}{4}\left[U_{[F]}^{intra}(x,x')(1+r_{1}r_{2}r_{3}r_{4})\right.\\
+\left.U_{[F]}^{inter}(x,x')(r_{2}r_{3}+r_{1}r_{4})\right],\label{eq:U_rF_Uint}\end{multline}
where \begin{multline}
U_{[F]}^{intra/inter}(x,x')=L^{2}\int\int d^{2}r_{\perp}d^{2}r'_{\perp}\\
\times\varphi_{pF_{1}}^{*}(\vec{r})\varphi_{\pm pF_{2}}^{*}(\vec{r}')\varphi_{\pm pF_{3}}(\vec{r}')\varphi_{pF_{4}}(\vec{r})U(\vec{r}-\vec{r}').\label{eq:Uintra_INTER}\end{multline}
Note that the 3D extention of the considered SWNT enters the effective
1D interaction potential via equation (\ref{eq:Uintra_INTER}). In
Appendix \ref{sec:Modeling-the-interaction} we show how we actually
determine the values for the potentials $U_{[F]}^{intra/inter}(x,x')$.

\subsubsection{The relevant scattering processes\label{sub:The-relevant-scattering}}

\begin{figure}
\begin{center}\includegraphics[%
  width=0.90\columnwidth]{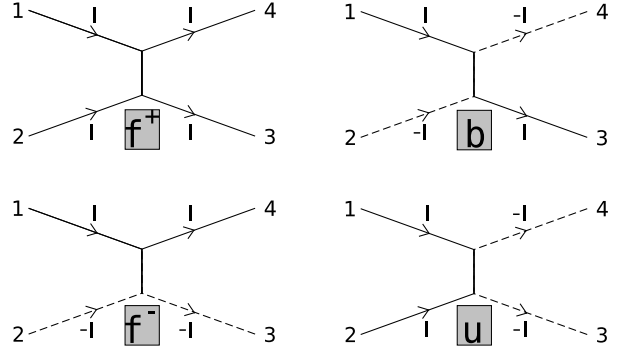}\end{center}

\caption{\label{cap:The-relevant-scattering}The relevant scattering processes.
Forward/back/Umklapp scattering are denoted by $f^{\pm}/b/u$. The
index I represents one of the three degrees of freedom $r,F,\sigma$
(branch, Fermi point and spin, respectively).}
\end{figure}

Not all of the terms in (\ref{eq:V1D_rs}) contribute to the interaction
because the corresponding potential $U_{[r][F]}$ vanishes or has
a very small amplitude. In order to pick out the relevant terms, it
is convenient to introduce the notion of forward ($f$)-, back ($b$)-
and Umklapp ($u$)- scattering with respect to an arbitrary index
quadruple $[I]$ associated to the electron operators in (\ref{eq:V1D_rs}).
Denoting the scattering type by $S_{I}$ we write $[I]_{S_{I}=f^{\pm}}$
for $[I,\pm I,\pm I,I]$. Furthermore we use $[I]_{S_{I}=b}$ for
$[I,-I,I,-I]$ and $[I]_{S_{I}=u}$ is equivalent to $[I,I,-I,-I]$,
cf. Fig. \ref{cap:The-relevant-scattering}. %
Keeping only the relevant terms, the interaction part of the Hamiltonian
acquires the form, \begin{equation}
V=\sum_{S_{r}=f,b,u}\sum_{S_{F}=r,b}\sum_{S_{\sigma}=f}V_{S_{r}S_{F}S_{\sigma}},\label{eq:V_scatt_types}\end{equation}
where \begin{multline}
V_{S_{r}S_{F}S_{\sigma}}:=\frac{1}{2}\sum_{\left\{ [r]_{S_{r}},[F]_{S_{F}},[\sigma]_{S_{\sigma}}\right\} }\int\int dx\, dx'U_{[r][F]}(x,x')\\
\times\psi_{r_{1}F_{1}\sigma}^{\dagger}(x)\psi_{r_{2}F_{2}\sigma'}^{\dagger}(x')\psi_{r_{3}F_{3}\sigma'}(x')\psi_{r_{4}F_{4}\sigma}(x),\label{eq:nonrhorho1}\end{multline}
as we are going to demonstrate in the following.

\paragraph{Scattering of $r$\label{sub:Scatt_r}}

We start with the possible scattering events related to the pseudo
spin $r$. From (\ref{eq:U_rF_Uint}) we can immediately read off
that the interaction potential $U_{[r][F]}$ does not vanish only
if $r_{2}r_{3}=r_{1}r_{4}.$ Thus we find the following cases for
the relevant scattering types, \[
i)\: r_{1}=r_{4},\, r_{2}=r_{3}\textrm{ and }ii)\: r_{1}=-r_{4},\, r_{2}=-r_{3}.\]
 Relation $i)$ summarizes all the forward scattering processes with
respect to $r$ and the associated interaction potential is, \begin{multline}
U_{[r]_{f}[F]}(x,x')=\frac{1}{2}\left[U_{[F]}^{intra}(x,x')+U_{[F]}^{inter}(x,x')\right]\\
=:U_{[F]}^{+}(x,x').\label{eq:U+}\end{multline}
 Case $ii)$ includes all $S_{r}=b$ and $S_{r}=u$ processes and
here the interaction potential is proportional to the difference between
$U^{intra}$ and $U^{inter}$,\begin{multline}
U_{[r]_{b/u}[F]}(x,x')=\frac{1}{2}\left[U_{[F]}^{intra}(x,x')-U_{[F]}^{inter}(x,x')\right]\\
=:U_{[F]}^{\Delta}(x,x').\label{eq:Udelta}\end{multline}

\paragraph{Scattering of $F$\label{sub:Scatt_F}}

The determination of the essential scattering processes with respect
to $F$ can be achieved by exploiting the approximate conservation
of quasi momentum. Looking at expression (\ref{eq:phi_pF_chi}) for
the wave functions $\varphi_{pF}(\vec{r})$, we find that the interaction
potential $U_{[r][F]}$, cf. (\ref{eq:U_rF_origin}), contains phase
factors of the form $e^{-i\left(F_{1}-F_{4}\right)R_{x}}\times$ $e^{-i\left(F_{2}-F_{3}\right)R'_{x}}$.
Although we are considering a finite system, therefore not having
perfect translational symmetry, after the integration along the tube
axis in (\ref{eq:V1D_rs}), only terms without fast oscillations survive
\footnote{For a perfectly translational invariant 1D system it holds 

\begin{multline*}
\int_{0}^{L}dx'\int_{0}^{L}dx\, U(x-x')e^{ikx}e^{ik'x'}= \\ 
\int_{0}^{L}dx'\int_{-x'}^{L-x'}dy\, U(y)e^{iky}e^{i(k+k')x'}=\tilde{U}_{k}\int_{0}^{L}dx'e^{i(k+k')x'},
\end{multline*}
where $\tilde{U}_{k}=\int_{-x'}^{L-x'}dyU(y)e^{iky}$ does not depend
on $x'$ because we have assumed translational invariance. So it is
clear that the double integral vanishes unless $k+k'\approx0.$ %
}. The corresponding condition is given by \begin{equation}
F_{1}-F_{4}+F_{2}-F_{3}=0,\label{eq:condonF}\end{equation}
 that means only the $S_{F}=f$ and $S_{F}=b$ terms survive. We have
explicitly checked that due to the discrete nature of the SWNT lattice
also the $S_{F}=u$ processes have very small amplitudes and can be
neglected. Note that condition (\ref{eq:condonF}) leads to $\mathrm{sgn}(F_{1}F_{2}F_{3}F_{4})=1$
in (\ref{eq:V1D_rs}).

\paragraph{Scattering of $\sigma$\label{sub:Scatt_s}}

It is clear that only $S_{\sigma}=f$ processes are allowed, since
the Coulomb interaction is spin independent.

Altogether this proofs equation (\ref{eq:V_scatt_types}).

\paragraph{Processes conserving or not conserving the fermionic configuration }

From the discussion in Section \ref{sec:The-noninteracting-System}
we already know that we have to distinguish between electrons with
different spin $\sigma$ and pseudo spin $r$. In the following we
will denote the number of electrons of a certain species by $N_{r\sigma}$
and we will refer to the quantity $\vec{N}=(N_{+\uparrow},N_{+\downarrow},N_{-\uparrow},N_{-\downarrow})$
as fermionic configuration. Not all of the scattering processes in
(\ref{eq:V_scatt_types}) conserve $\vec{N}.$ In more detail, for
terms with $(S_{r},S_{\sigma})=(u,f^{+})$, $(S_{r},S_{\sigma})=(b,f^{-})$
and $(S_{r},S_{\sigma})=(u,f^{-})$ $\vec{N}$ is not a good quantum
number as can be easily verified by using equation (\ref{eq:nonrhorho1}).
In general, only processes described by the $\vec{N}$ conserving
terms are sensitive to the total number of electrons in the dot. As
example we mention the charging energy contribution proportional to
$N_{c}^{2}$, $N_{c}:=\sum_{r\sigma}N_{r\sigma}$ arising from the
$(S_{r},S_{F},S_{\sigma})=(f,f,f)$ processes appearing later on.
On the other hand for the $\vec{N}$ non conserving terms, only the
vicinity of the Fermi surface is of relevance.

\paragraph{Processes only relevant near half-filling}

Away from half-filling we find that terms with \begin{equation}
r_{1}F_{1}+r_{2}F_{2}-r_{3}F_{3}-r_{4}F_{4}\neq0,\label{eq:Condition_hf}\end{equation}
i.e., the Umklapp scattering terms with respect to the product $rF$
\footnote{There are simple rules for determining the scattering type $S_{rF}$
if $S_{r}$ and $S_{F}$ are known. Defining a product by $S_{rF}=S_{r}S_{F}=S_{F}S_{r}$
it holds, $Sf^{+}=S;$ $S^{2}=f^{+};$ $f^{-}u=b;$ $f^{-}b=u$ and
$ub=f^{-}.$ %
} can be neglected in (\ref{eq:V1D_rs}). For the $\vec{N}$ non conserving
terms fulfilling (\ref{eq:Condition_hf}) this is a consequence of
the approximate conservation of quasi momentum, arising from the slow
oscillations of the 1D electron operators in (\ref{eq:nonrhorho1})
which near the Fermi surface are given by the exponential $e^{-i\left[\left(r_{1}F_{1}N_{r_{1}\sigma_{1}}-r_{4}F_{4}N_{r_{4}\sigma_{4}}\right)x+\left(r_{2}F_{2}N_{r_{2}\sigma_{2}}-r_{3}F_{3}N_{r_{3}\sigma_{3}}\right)x'\right]}$.
After performing the integrations in (\ref{eq:nonrhorho1}) this leads
approximately to (\ref{eq:Condition_hf}). The $\vec{N}$ conserving
terms obeying (\ref{eq:Condition_hf}), $V_{f^{-}bf}$ and $V_{bf^{-}f^{+}}$,
which describe not only processes near the Fermi level, add a term
proportional to the number of electrons above half-filling to the
Hamiltonian, therefore just giving rise to a shift of the chemical
potential.

\subsubsection{Long ranged vs. short ranged interactions}

Except of $U_{[r]_{f}[F]_{f}}=U_{[F]_{f}}^{+}$, all relevant interaction
potentials $U_{[r][F]}$ can effectively be treated as local interactions:
In the case of $U_{[F]_{b}}^{+}$ this is due to the appearance of
phase factors $e^{i2F(R_{x}-R'_{x})}$ in (\ref{eq:Uintra_INTER}),
arising from the Bloch waves $\varphi_{pF}(\vec{r})$, cf. equation
(\ref{eq:phi_pF_chi}), oscillating much faster than the electron
operators $\psi_{r\sigma F}(x).$ The potentials $U_{[F]}^{\Delta}$,
being proportional to the difference of the inter- and intra-lattice
interaction potentials, are in general short ranged, since $U_{[F]}^{intra}(x,x')$
and $U_{[F]}^{inter}(x,x')$ only have considerably differing values
for $\left|x-x'\right|\lesssim a_{0}$ with the next neighbour distance
$a_{0}=0.142$ nm of the carbon atoms in the SWNT lattice \cite{Egger1997}.
Summarizing, \textit{only the processes with $(S_{r},S_{F})=(f,f)$
are long ranged}. All other terms can effectively be written as local
interactions. I.e. for $(S_{r},S_{F})\neq(f,f)$ we can use the approximation
\begin{equation}
\frac{1}{2}U_{[r]_{S_{r}}[F]_{S_{F}}}(x,x')\approx Lu_{S_{r}\, S_{F}}\delta(x-x'),\label{eq:Def_uSrSF}\end{equation}
where we have introduced the coupling parameters \begin{equation}
u^{S_{r}\, S_{F}}:=1/(2L^{2})\int\int dx\, dx'U_{[r]_{S_{r}}[F]_{S_{F}}}(x,x').\label{eq:couplingparameters}\end{equation}
Using the approximation (\ref{eq:Def_uSrSF}) we obtain from (\ref{eq:nonrhorho1})
in the case $(S_{r},S_{F})\neq(f,f)$ the following expression for
the non forward scattering interaction terms,\begin{multline}
V_{S_{r}S_{F}S_{\sigma}}\approx Lu_{S_{r}\, S_{F}}\sum_{\left\{ [r]_{S_{r}},[F]_{S_{F}},[\sigma]_{f}\right\} }\\
\times\int_{0}^{L}dx\psi_{r_{1}F_{1}\sigma}^{\dagger}(x)\psi_{r_{2}F_{2}\sigma'}^{\dagger}(x)\psi_{r_{3}F_{3}\sigma'}(x)\psi_{r_{4}F_{4}\sigma}(x).\label{eq:nonrhorhob}\end{multline}
In the following we use the abbreviations $u^{+}:=u_{f\, b}$ and
$u^{\Delta}_{S_{F}}:=u_{b\, S_{F}}=u_{u\, S_{F}}.$ For details about
the calculation, see Appendix \ref{sec:Modeling-the-interaction}.%
\begin{table}
\begin{center}\renewcommand\arraystretch{1.7}\begin{tabular}{|c|c|c|c|}
\hline 
&
$\,\frac{u^{+}d}{\varepsilon_{0}}\,$&
$\,\frac{u^{\Delta}_{f}d}{\varepsilon_{0}}\,$&
$\,\frac{u^{\Delta}_{b}d}{\varepsilon_{0}}\,$\tabularnewline
\hline 
$\epsilon=1.4$&
$0.22\textrm{\AA{}}$&
$0.14\textrm{\AA{}}$&
$0.22\textrm{\AA{}}$\tabularnewline
\hline 
$\epsilon=2.4$&
$0.28\textrm{\AA{}}$&
$0.22\textrm{\AA{}}$&
$0.28\textrm{\AA{}}$\tabularnewline
\hline
\end{tabular}\end{center}

\caption{\label{cap:us}The dependence of the coupling constants $u^{+},$
$u^{\Delta}_{f}$ and $u^{\Delta}_{b}$ on the tube diameter $d$
and on the dielectric constant $\epsilon$. }
\end{table}
We find that in general the coupling constants $u^{+}$ and $u^{\Delta}_{S_{F}}$
scale inversely with the total number of lattice sites, i.e., like
$1/Ld$, where $d$ is the tube diameter. From a physical point of
view this is due to an increasing attenuation of the wave functions
for a growing system size. Therefore the probability of processes
mediated by local interactions is proportional to $1/Ld$. Because
the level spacing of the noninteracting system $\varepsilon_{0}$
scales like $1/L$, cf. (\ref{eq:levelspacing_nonint}), the products
$u^{+}d/\varepsilon_{0}$ and $u^{\Delta}_{S_{F}}d/\varepsilon_{0}$
are constants. The corresponding numerical values for different dielectric
constants $\epsilon$, cf. equation (\ref{eq:Ohno}), are given in
table \ref{cap:us}.

\subsubsection{Density-density vs. non-density-density processes}

The interaction processes can be divided into density-density terms,
easily diagonalizable by bosonization \cite{Delft1998}, and non-density-density terms
respectively. It is clear that the forward scattering interaction
$V_{f\, f\, f}$ is of density-density form, \begin{multline}
V_{f\, f\, f}=\frac{1}{2}\sum_{rr'}\sum_{FF'}\sum_{\sigma\sigma'}\int\int dx\, dx'U_{[F]_{f}}^{+}(x,x')\\
\times\rho_{rF\sigma}(x)\rho_{r'F'\sigma'}(x'),\label{eq:VFSFSFS}\end{multline}
where the densities $\rho_{rF\sigma}(x)$ are given by \[
\rho_{rF\sigma}(x)=\psi_{rF\sigma}^{\dagger}(x)\psi_{rF\sigma}(x).\]
But since we treat the short ranged interactions as local, also $V_{f^{+}\, b\, f^{+}}$,
\begin{multline}
V_{f^{+}\, b\, f^{+}}=\\
Lu^{+}\sum_{r\sigma F}\int_{0}^{L}dx\psi_{rF\sigma}^{\dagger}(x)\psi_{r-F\sigma}^{\dagger}(x)\psi_{rF\sigma}(x)\psi_{r-F\sigma}(x)\\
=-Lu^{+}\sum_{r\sigma F}\int_{0}^{L}dx\rho_{rF\sigma}(x)\rho_{r-F\sigma}(x),\label{eq:VFSBSFS}\end{multline}
and similarly $V_{b\, f^{+}/b\, f^{+}},$\begin{multline}
V_{b\, f^{+}/b\, f^{+}}=\\
-Lu^{\Delta}_{f^{+}/b}\sum_{r\sigma F}\int_{0}^{L}dx\rho_{rF\sigma}(x)\rho_{-r\pm F\sigma}(x),\label{eq:VBSFSBSFS}\end{multline}
are density-density interactions. In total the density-density part
of the interaction is given by \begin{equation}
V_{\rho\rho}=V_{f\, f\, f}+V_{f^{+}\, b\, f^{+}}+V_{b\, f^{+}\, f^{+}}+V_{b\, b\, f^{+}}.\label{eq:Vrr_expl}\end{equation}
The remaining terms are not of density-density form and are collected
in the operator $V_{\mathrm{n}\rho\rho}.$ Including only the contributions
relevant away from half-filling, we obtain,\begin{equation}
V_{\mathrm{n}\rho\rho}=V_{f^{+}\, b\, f^{-}}+V_{b\, f^{+}\, f^{-}}+V_{b\, b\, f^{-}}+V_{u\, f^{-}\, f}+V_{u\, b\, f}.\label{eq:Vnrr_expl1}\end{equation}
Near half-filling additionally the processes\begin{equation}
V_{f^{-}\, b\, f},V_{b\, f^{-}\, f}\textrm{ and }V_{u\, f^{+}\, f^{-}},\label{eq:Vnrr_explhf}\end{equation}
satisfying condition (\ref{eq:Condition_hf}), contribute to $V_{n\rho\rho}$.
Overall, the SWNT Hamiltonian acquires the form, \[
H=H_{0}+V_{\rho\rho}+V_{\mathrm{n}\rho\rho}.\]

\section{Expressing the SWNT Hamiltonian in the eigenbasis of $H_{0}+V_{\rho\rho}$\label{sec:The-SWNT-Spectrum}}

Away from half-filling, the interaction is dominated by $V_{f\, f\, f}$.
Together with $H_{0}$ it yields the {}``standard'' theory for interacting
electrons in SWNTs \cite{Egger1997,Odintsov1999,Kane1997}. Using
bosonization we will in the next step diagonalize $H_{0}+V_{\rho\rho}$.
Subsequently we will examine the effect of $V_{\mathrm{n}\rho\rho}$
by calculating the matrix elements of $V_{\mathrm{n}\rho\rho}$ between
the eigenstates of $H_{0}+V_{\rho\rho}$. The diagonalization of $V_{\mathrm{n}\rho\rho}$
in a truncated eigenbasis of $H_{0}+V_{\rho\rho}$, discussed in Section
\ref{sec:The-SWNT-spectrumresults} then yields to a good approximation
the correct eigenstates and the spectrum of the total Hamiltonian
$H$.

\subsection{Diagonalizing $H_{0}+V_{\rho\rho}$\label{sub:Diagonalizing} }

By introducing operators creating/annihilating bosonic excitations
we can easily diagonalize $H_{0}+V_{\rho\rho}$ as we show in this
section. It turns out that the Fourier coefficients of the density
operators $\rho_{r\sigma F}(x)$ are essentially of bosonic nature.
In detail, we get by Fourier expansion, \begin{equation}
\rho_{rF\sigma}(x)=\frac{1}{2L}\sum_{q}e^{i\mathrm{sgn}(F)qx}\rho_{r\sigma q},\label{eq:FE}\end{equation}
where $q=\frac{\pi}{L}n_{q},\, n_{q}\in\mathbb{Z}.$ Then the operators
$b_{\sigma q_{r}}$ defined by,\begin{equation}
b_{\sigma q_{r}}:=\frac{1}{\sqrt{n_{q}}}\rho_{r\sigma q_{r}},\quad q_{r}:=r\cdot q,\quad q>0\label{eq:bop}\end{equation}
fulfill the canonical commutation relations $[b_{\sigma q},b_{\sigma'q'}^{\dagger}]=\delta_{\sigma'\sigma}\delta_{qq'}$
as shown e.g. in \cite{Delft1998}. For completeness we give the explicit
expression for $b_{\sigma q_{r}},\, r=\pm$, \[
b_{\sigma q_{r}}=\frac{1}{\sqrt{n_{q}}}\sum_{\kappa}c_{r\sigma\kappa}^{\dagger}c_{r\sigma\kappa+q_{r}},\quad q>0.\]
 The bosonized expression for $H_{0}$ is well known \cite{Mayrhofer2006},
\begin{equation}
H_{0}=\sum_{r\sigma}\left[\varepsilon_{0}\sum_{q>0}\left|n_{q}\right|b_{\sigma q_{r}}^{\dagger}b_{\sigma q_{r}}+\frac{\varepsilon_{0}}{2}\mathcal{N}_{r\sigma}^{2}+r\frac{\varepsilon_{\Delta}}{2}\mathcal{N}_{r\sigma}\right],\label{eq:H_0_bos}\end{equation}
Here the first term describes collective particle-hole excitations,
whereas the second term is due to Pauli's principle and represents
the energy cost for the shell filling. The third term accounts for
a possible energy mismatch between the bands $r=\pm$, given by \[
\varepsilon_{\Delta}=\mathrm{sgn}(\Delta)\varepsilon_{0}\min(2\left|\Delta\right|,|2\left|\Delta\right|-1|).\]
 The operators $\mathcal{N}_{r\sigma}$ count the number of electrons
$N_{r\sigma}$ in branch $(r\sigma).$ Bosonization of $V_{\rho\rho}$
can be achieved by inserting the Fourier expansion (\ref{eq:FE}) into
expressions (\ref{eq:VFSFSFS}), (\ref{eq:VFSBSFS}) and (\ref{eq:VBSFSBSFS}),
thereby making use of definition (\ref{eq:bop}). We obtain,\begin{multline}
V_{\rho\rho}=V_{f\, f\, f}+V_{f^{+}\, b\, f^{+}}+V_{b\, f^{+}/b\, f^{+}}=\\
\frac{1}{2}\sum_{q>0}n_{q}\left\{ W_{q}\left[\sum_{r\sigma}\left(b_{\sigma r\cdot q}+b_{\sigma r\cdot q}^{\dagger}\right)\right]^{2}\right.\\
-u^{+}\sum_{r\sigma}\left(b_{\sigma r\cdot q}b_{\sigma r\cdot q}+h.c.\right)\\
-u^{\Delta}_{f}\sum_{r\sigma}\left(b_{\sigma r\cdot q}b_{\sigma-r\cdot q}+h.c.\right)\\
\left.-u^{\Delta}_{b}\sum_{r\sigma}\left(b_{\sigma r\cdot q}b_{\sigma-r\cdot q}^{\dagger}+h.c.\right)\right\} \\
+\frac{1}{2}\left[E_{c}\mathcal{N}_{c}^{2}-\frac{J}{2}\sum_{r\sigma}\mathcal{N}_{r\sigma}\mathcal{N}_{-r\sigma}-u^{+}\sum_{r\sigma}\mathcal{N}_{r\sigma}^{2}\right],\label{eq:V_rr_bos}\end{multline}
where the coefficients $W_{q}$ determine the interaction strength
of $V_{f\, f\, f}$ and are given by \[
W_{q}=\frac{1}{L^{2}}\int dx\int dx'U_{[F]_{f}}^{+}(x,x')\cos(qx)\cos(qx').\]
The last line of (\ref{eq:V_rr_bos}) describes the contribution of
$V_{\rho\rho}$ to the system energy depending on the number of electrons
in the single branches $(r\sigma)$. Here $E_{c}=W_{0}$ is the SWNT
charging energy, $\mathcal{N}_{c}=\sum_{r\sigma}\mathcal{N}_{r\sigma}$
counts the total number of electrons. Spin alignment of electrons
with different branch index $r$ is favoured by the term proportional
to $J/2:=u^{\Delta}_{f}+u^{\Delta}_{b}$. Finally the term coupling with
$u^{+}$ counteracts the energy cost for the shell filling in equation
(\ref{eq:H_0_bos}).

Since the bosonic operators appear quadratically in (\ref{eq:H_0_bos})
and (\ref{eq:V_rr_bos}) we can diagonalize $H_{0}+V_{\rho\rho}$
by introducing new bosonic operators $a_{j\delta q}$ and $a_{j\delta q}^{\dagger}$
via the Bogoliubov transformation \cite{Avery1976} given below by equation (\ref{eq:b_a}).
We obtain\begin{multline}
H_{0}+V_{\rho\rho}=\sum_{j\delta}\sum_{q>0}\varepsilon_{j\delta q}a_{j\delta q}^{\dagger}a_{j\delta q}+\frac{1}{2}E_{c}\mathcal{N}_{c}^{2}\\
+\frac{1}{2}\sum_{r\sigma}\mathcal{N}_{r\sigma}\left[-\frac{J}{2}\mathcal{N}_{-r\sigma}+\left(\varepsilon_{0}-u^{+}\right)\mathcal{N}_{r\sigma}+r\varepsilon_{\Delta}\right].\label{eq:H0Vrr_diag}\end{multline}
The first term describes the bosonic excitations of the system, created/annihilated
by the operators $a_{j\delta q}^{\dagger}$ / $a_{j\delta q}$. The
four channels $j\delta=c+,c-,s+,s-$ are associated to total $(+)$
and relative $(-)$ (with respect to the index $r$) spin $(s)$ and
charge $(c)$ excitations. The decoupling of the four modes $j\delta$,
the so called spin-charge separation, will be partly broken by $V_{\mathrm{n}\rho\rho}$.
The excitation energies $\varepsilon_{j\delta q}$ and the relation
between the new bosonic operators $a_{j\delta q}$ and the old operators
$b_{\sigma q_{r}}$ are determined by the Bogoliubov transformation.
In detail, we find with $\varepsilon_{0q}:=\varepsilon_{0}n_{q},$
\[
\varepsilon_{c+q}=\varepsilon_{0q}\sqrt{1+8W_{q}/\varepsilon_{0}},\]
\[
\varepsilon_{s/c-q}=\varepsilon_{0q}(1-u^{\Delta}_{b}/\varepsilon_{0})\]
 and \[
\varepsilon_{s+q}=\varepsilon_{0q}(1+u^{\Delta}_{b}/\varepsilon_{0}).\]
 The energies of the $c+$ channel are largely enhanced compared to
the other excitations because of the dominating $V_{f\, f\, f}$ contribution.
For small $q$ the ratio $g_{q}:=\varepsilon_{0q}/\varepsilon_{c+q}$
is approximately $0.2$, whereas for large $q$ it tends to $1$ \cite{Mayrhofer2006}.
Small corrections due to the coupling constants $u^{\Delta}_{f}$
and $u^{+}$ have been neglected. For the transformation from the
old bosonic operators $b_{\sigma q_{r}}$ to the new ones $a_{j\delta q}$
we find \begin{equation}
b_{\sigma q_{r}}=\sum_{j\delta}\Lambda_{r\sigma}^{j\delta}\left(B_{j\delta q}a_{j\delta q}+D_{j\delta q}a_{j\delta q}^{\dagger}\right),\quad q>0\label{eq:b_a}\end{equation}
where \begin{equation}
\Lambda_{r\sigma}^{j\delta}=\frac{1}{2}\left(\begin{array}{cccc}
1 & 1 & 1 & 1\\
1 & 1 & -1 & -1\\
1 & -1 & 1 & -1\\
1 & -1 & -1 & 1\end{array}\right),\quad\begin{array}{c}
j\delta=c+,c-,s+,s-\\
r\sigma=+\uparrow,+\downarrow,-\uparrow,-\downarrow\end{array}.\label{eq:Lambdajd_rs}\end{equation}
The transformation coefficients $B_{j\delta q}$ and $D_{j\delta q}$
in the case of the three modes $j\delta=c-,s+,s-$ are given by \begin{equation}
B_{j\delta q}=1\textrm{ and }D_{j\delta q}=0\label{eq:SC_ntrl}\end{equation}
and for $j\delta=c+$ we obtain \begin{equation}
B_{j\delta q}=\frac{1}{2}\left(\sqrt{g_{q}}+\frac{1}{\sqrt{g_{q}}}\right),\, D_{j\delta q}=\frac{1}{2}\left(\sqrt{g_{q}}-\frac{1}{\sqrt{g_{q}}}\right),\label{eq:SCcp}\end{equation}
with $g_{q}=\frac{\varepsilon_{0q}}{\varepsilon_{c+q}}$. Small corrections
to (\ref{eq:SC_ntrl}) and (\ref{eq:SCcp}) resulting from the terms
$V_{f^{+}\, b\, f^{+}}$ and $V_{b\, f^{+}/b\, f^{+}}$ have been
neglected. 

The physical meaning of the fermionic contributions in (\ref{eq:H0Vrr_diag}),
depending on the number counting operators, have already been discussed
subsequently to equations (\ref{eq:H_0_bos}) and (\ref{eq:V_rr_bos})
respectively. 

An eigenbasis of $H_{0}+V_{\rho\rho}$ is formed by the states \begin{equation}
\left|\vec{N},\vec{m}\right\rangle :=\prod_{j\delta q}\frac{\left(a_{j\delta q}^{\dagger}\right)^{m_{j\delta q}}}{\sqrt{m_{j\delta q}!}}\left|\vec{N},0\right\rangle ,\label{eq:eigenstates_H0Vrr}\end{equation}
 where $\left|\vec{N},0\right\rangle $ has no bosonic excitation.
Remember that the fermionic configuration $\vec{N}=(N_{-\uparrow},N_{-\downarrow},N_{+\uparrow},N_{+\downarrow})$
defines the number of electrons in each of the branches $(r\sigma)$.
In the following we will use the states from (\ref{eq:eigenstates_H0Vrr})
as basis to examine the effect of $V_{\mathrm{n}\rho\rho}.$ For this
purpose we evaluate in the next section the corresponding matrix elements
using the bosonization identity for the 1D electron operators.

\subsection{The matrix elements $\left\langle \vec{N}\vec{m}\left|V_{\mathrm{n}\rho\rho}\right|\vec{N}'\vec{m}'\right\rangle $\label{sub:The-matrix-elements}}

Generally, due to $V_{\mathrm{n}\rho\rho}$, the quantities $\vec{N}$
and $\vec{m}$ are not conserved. Especially, the terms with $S_{r}=b,\, u$
in (\ref{eq:Vnrr_expl1}) mix states with different $\vec{N}$. However,
denoting \[
N_{s}:=\sum_{r\sigma}\mathrm{sgn}(\sigma)N_{r\sigma},\]
 \[
N_{c}^{-}:=\sum_{r\sigma}\mathrm{sgn}(r)N_{r\sigma}\]
 and \[
N_{s}^{-}:=\sum_{r\sigma}\mathrm{sgn}(r\sigma)N_{r\sigma}\]
 we find that ($N_{c},\, N_{s},\, N_{c}^{-}\textrm{ mod }4\,,N_{s}^{-}\textrm{ mod }4$)
is conserved, i.e., states differing in those quantities do not mix,
such that the corresponding matrix elements of $V_{\mathrm{n}\rho\rho}$
are zero. Note that in contrast to the real spin $S_{z}=\frac{1}{2}N_{s}$,
the pseudo spin $\tilde{S}_{z}=\frac{1}{2}N_{c}^{-}$ is not conserved
in general. 

We already know that all the processes $V_{S_{r}S_{F}S_{\sigma}}$
contained in $V_{\mathrm{n}\rho\rho}$ are effectively local interactions,
i.e., of the form (\ref{eq:nonrhorhob}). Hence, in order to calculate
the corresponding matrix elements $\left\langle \vec{N}\vec{m}\left|V_{S_{r}S_{F}S_{\sigma}}\right|\vec{N}\vec{'m}'\right\rangle $
we first derive an expression for \begin{multline}
M_{[r][F][\sigma]}(\vec{N},\vec{m},\vec{N}',\vec{m}',x):=\\
\left\langle \vec{N}\vec{m}\left|\psi_{r_{1}\sigma F_{1}}^{\dagger}(x)\psi_{r_{2}\sigma'F_{2}}^{\dagger}(x)\psi_{r_{3}\sigma'F_{3}}(x)\psi_{r_{4}\sigma F_{4}}(x)\right|\vec{N}\vec{'m}'\right\rangle .\label{eq:DefMNmNmx}\end{multline}
For this purpose we express the operators $\psi_{r\sigma F}(x)$ in
terms of the bosonic operators $b_{\sigma r\cdot q}$ and $b_{\sigma r\cdot q}^{\dagger},\, q>0$,
using the bosonization identity \cite{Delft1998}, \begin{equation}
\psi_{r\sigma F}(x)=\eta_{r\sigma}K_{r\sigma F}(x)e^{i\phi_{r\sigma F}^{\dagger}(x)}e^{i\phi_{r\sigma F}(x)}.\label{eq:bosident}\end{equation}
The operator $\eta_{r\sigma}$ is the so called Klein factor, which
annihilates an electron in the ($r\sigma$) branch and thereby takes
care of the right sign as required from the fermionic anticommutation
relations, in detail, \begin{equation}
\eta_{r\sigma}\left|\vec{N},\vec{m}\right\rangle =(-1)^{\sum_{l=1}^{(r\sigma)-1}N_{l}}\left|\vec{N}-\hat{e}_{r\sigma},\vec{m}\right\rangle ,\label{eq:Def_Klein_factors}\end{equation}
 where we use the convention $l=+\uparrow,+\downarrow,-\uparrow,-\downarrow=1,2,3,4$.
$K_{r\sigma F}(x)$ yields a phase factor depending on the number
of electrons in $(r\sigma),$\begin{equation}
K_{r\sigma F}(x)=\frac{1}{\sqrt{2L}}e^{i\frac{\pi}{L}\mathrm{sgn}(F)(r\cdot\mathcal{N}_{r\sigma}+\Delta)x}.\label{eq:K_rsF}\end{equation}
Finally, we have the boson fields $i\phi_{r\sigma F}(x)$, \begin{equation}
i\phi_{r\sigma F}(x)=\sum_{q>0}\frac{1}{\sqrt{n_{q}}}e^{i\mathrm{sgn}(rF)qx}b_{\sigma r\cdot q}.\label{eq:phifield_b}\end{equation}
 In Appendix \ref{sec:CalcMmm} we are going to demonstrate with the help of the
bosonization identity (\ref{eq:bosident}), that the matrix elements
from equation (\ref{eq:DefMNmNmx}) factorize into a fermionic and
a bosonic part,\begin{multline*}
M_{[r][F][\sigma]}(\vec{N},\vec{m},\vec{N}',\vec{m}',x)=\\
M_{[r][F][\sigma]}(\vec{N},\vec{N}',x)M_{[r][F][\sigma]}(\vec{m},\vec{m}',x),\end{multline*}
where the fermionic part is given by

\begin{multline}
M_{[l]}(\vec{N},\vec{N}',x)=\\
\left\langle \vec{N}\right|K_{l_{1}}^{\dagger}(x)\eta_{l_{1}}^{\dagger}K_{l_{2}}^{\dagger}(x)\eta_{l_{2}}^{\dagger}K_{l_{3}}(x)\eta_{l_{3}}K_{l_{4}}(x)\eta_{l_{4}}\left|\vec{N}'\right\rangle \end{multline}
and the bosonic part reads\begin{multline}
M_{[l]}(\vec{m},\vec{m}',x)=\left\langle \vec{m}\right|e^{-i\phi_{l_{1}}^{\dagger}(x)}e^{-i\phi_{l_{1}}(x)}e^{-i\phi_{l_{2}}^{\dagger}(x)}e^{-i\phi_{l_{2}}(x)}\\
e^{i\phi_{l_{3}}^{\dagger}(x)}e^{i\phi_{l_{3}}(x)}e^{i\phi_{l_{4}}^{\dagger}(x)}e^{i\phi_{l_{4}}(x)}\left|\vec{m}'\right\rangle .\label{eq:Mmm_def_mt}\end{multline}
In order to improve readability we have replaced the indices $rF\sigma$
by a single index $l.$ As we demonstrate in Appendix \ref{sec:CalcMmm},
the explicit evaluation yields\begin{multline}
M_{[r][F][\sigma]}(\vec{N},\vec{N}',x)=\\
\frac{1}{(2L)^{2}}\delta_{\vec{N},\vec{N}'+\vec{E}_{[r][\sigma]}}T_{\vec{N}\vec{N}'[r][\sigma]}Q_{\vec{N}\vec{N}'[r][F]}(x),\label{eq:MNN_TQ}\end{multline}
where $\vec{E}_{[r][\sigma]}:=\vec{e}_{r_{1}\sigma}+\vec{e}_{r_{2}\sigma'}-\vec{e}_{r_{3}\sigma'}-\vec{e}_{r_{4}\sigma}.$
The Klein factors in (\ref{eq:bosident}) lead to the sign factor
$T_{\vec{N}\vec{N}'[r][\sigma]}$ which is either $+1$ or $-1$ and
$Q_{\vec{N}\vec{N}'[r][F]}(x)$ yields a phase depending on $\vec{N}$.
Explicit expressions can be found in Appendix \ref{sec:CalcMmm},
equations (\ref{eq:TNNp}) to (\ref{eq:Q_NrF}). 

For the bosonic part of $M_{[r][F][\sigma]}(\vec{N},\vec{m},\vec{N}',\vec{m}',x)$
the calculation in Appendix \ref{sec:CalcMmm} leads to \begin{multline}
M_{[r][F][\sigma]}(\vec{m},\vec{m}',x)=C_{[r][F][\sigma]}(x)\\
\times A_{S_{rF}}(x)\prod_{j\delta q}F(\tilde{\lambda}_{[r][F][\sigma]}^{j\delta q}(x),m_{j\delta q},m'_{j\delta q}).\label{eq:Mmm_CF}\end{multline}
 Here the function $F(\lambda,m,m')$ stems from the evaluation of
matrix elements of the form $\left\langle m\left|e^{-\lambda^{*}a^{\dagger}}e^{\lambda a}\right|m'\right\rangle $,
where the bosonic excitations $\left|m\right\rangle $ are created
by the operators $a^{\dagger}$, i.e., $\left|m\right\rangle =\left(a^{\dagger}\right)^{m}/\sqrt{m!}\left|0\right\rangle .$
For the explicit form of $F(\lambda,m,m'),$ and the coefficients
$\tilde{\lambda}_{[r][F][\sigma]}^{j\delta q}(x)$, see Appendix \ref{sec:CalcMmm}.
The function $C_{[r][F][\sigma]}(x)$ is conveniently considered in
combination with $Q_{\vec{N}\vec{N}'[r][F]}(x),$ namely the product
\[
\tilde{K}_{\vec{N}[r][F][\sigma]}(x):=Q_{\vec{N}\vec{N}'[r][F]}(x)C_{[r][F][\sigma]}(x)\]
can be reexpressed as \begin{equation}
\tilde{K}_{\vec{N}[r][F][\sigma]}(x)=\tilde{Q}_{\vec{N}[r][F]}(x)\tilde{C}_{S_{r}S_{F}S_{\sigma}}(x),\label{eq:Ktilde_QC}\end{equation}
where \begin{multline*}
\tilde{Q}_{\vec{N}[r][F]}(x)=\\
\exp\left[-i\frac{\pi}{L}\left(\tilde{\sum}_{j=1}^{4}\mathrm{sgn}(r_{j}F_{j})N_{r_{j}\sigma_{j}}+\sum_{j=3}^{4}\mathrm{sgn}(r_{j}F_{j})\right)x\right]\end{multline*}
Here $\tilde{\sum}_{l=1}^{4}a_{l}$ denotes the sum $a_{1}+a_{2}-a_{3}-a_{4}$.
For $\tilde{C}_{S_{r}S_{F}S_{\sigma}}(x)$ we obtain \begin{multline}
\tilde{C}_{f^{+}bf^{-}}(x)=-\tilde{C}_{f^{-}bf}(x)=-\tilde{C}_{bf^{-}f^{+}}(x)=\\
1/4\sin^{2}\left(\frac{\pi}{L}x\right),\label{eq:Ctilde1}\end{multline}
 \begin{equation}
\tilde{C}_{ubf^{+}}(x)=-\tilde{C}_{uf^{-}f^{+}}(x)=4\sin^{2}\left(\frac{\pi}{L}x\right)\label{eq:Ctilde2}\end{equation}
 and $\tilde{C}_{S_{r}S_{F}S_{\sigma}}(x)\equiv1$ for the remaining
processes of $V_{\textrm{n}\rho\rho}$. 

\begin{figure}
\begin{center}\includegraphics[%
  width=1.0\columnwidth]{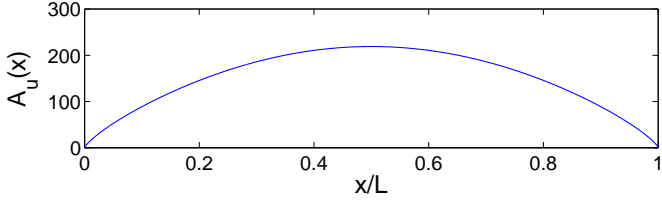}\end{center}

\caption{\label{cap:A_hf}$A_{S_{rF}=u}(x)$ as a function of $x$ for a (6,6)-SWNT.
Note the large magnitude of $A_{S_{rF}=u}(x)$ compared to $A_{S_{rF}\neq u}(x)\equiv1$
for the processes only relevant away from half-filling!}
\end{figure}

The function $A_{S_{rF}}(x)$ is differing from $1$ only for terms
with $S_{rF}=u$, i.e., for the terms fulfilling condition (\ref{eq:Condition_hf})
and which hence are relevant only near half-filling. The reason for
this is that only for the $S_{rF}=u$ terms the coefficients $\tilde{\lambda}_{[r][F][\sigma]}^{c+q}(x)$
related to the charged $c+$mode are not vanishing. Hence $A_{u}(x)$
depends strongly on the energy dispersion of the $c+$ mode and therefore
on the forward scattering part of the interaction, in detail\[
A_{u}(x)=\exp\left[2\sum_{q>0}\frac{1}{n_{q}}\left(1-\frac{\varepsilon_{0q}}{\varepsilon_{c+q}}\right)\sin^{2}(qx)\right].\]
 Since for the repulsive Coulomb interaction $\varepsilon_{0q}/\varepsilon_{c+q}<1$
holds, we find $A_{S_{rF}=u}(x)\ge1$. In Fig. \ref{cap:A_hf} we
show $A_{S_{rF}=u}(x)$ for a (6,6) SWNT. It is the large magnitude
of $A_{u}(x)$, that poses problems for properly treating the situation
at half-filling. Moreover we can expect that even for large diameter
tubes, interaction processes with $S_{rF}=u$ can not be neglected
near half-filling. %
 Altogether, we get with equations (\ref{eq:nonrhorhob}), (\ref{eq:MNN_TQ})
and (\ref{eq:Mmm_CF}) for the single contributions to $V_{\mathrm{n}\rho\rho}$,
\begin{multline}
\left\langle \vec{N}\vec{m}\left|V_{S_{r}S_{F}S_{\sigma}}\right|\vec{N}\vec{'m}'\right\rangle =\\
\times\frac{1}{4L}u^{S_{r}\, S_{F}}\sum_{\left\{ [r]_{S_{r}},[F]_{S_{F}},[\sigma]_{S_{\sigma}}\right\} }\delta_{\vec{N},\vec{N}'+\vec{E}_{[r]\sigma\sigma'}}T_{\vec{N}S_{r}S_{\sigma}}\\
\times\int dx\,\tilde{K}_{\vec{N}[r][F]}(x)A_{S_{rF}}(x)\prod_{j\delta q}F(\tilde{\lambda}_{[r][F][\sigma]}^{j\delta q}(x),m_{j\delta q},m'_{j\delta q}).\label{eq:ME_V_nrr_expl}\end{multline}
The evaluation of (\ref{eq:ME_V_nrr_expl}) causes no problems except
for the $\vec{N}$ conserving processes with $(S_{r},S_{F},S_{\sigma})=(f^{+},b,f^{-}),$
$(f^{-},b,f),$ $(b,f^{-},f^{+})$, since then we find $\tilde{K}_{\vec{N}[r][F][\sigma]}\sim1/4\sin^{2}(\frac{\pi}{L}x)$,
cf. equations (\ref{eq:Ktilde_QC}) and (\ref{eq:Ctilde1}), causing the
integral in (\ref{eq:ME_V_nrr_expl}) to diverge for $\sum_{j\delta q}\left|m_{j\delta q}-m'_{j\delta q}\right|\le1$,
such that the evaluation of the corresponding matrix elements needs
special care in this case. The origin of this divergence lies in the
fact, that if no bosonic excitations are present, the $\vec{N}$ conserving
processes depend on the total number of electrons in the single branches
(compare to the fermionic contributions to $H_{0}+V_{\rho\rho}$ in
(\ref{eq:H0Vrr_diag})). Since the bosonization approach requires
the assumption of an infinitely deep Fermi sea \cite{Delft1998} this
leads, without the correct regularization, necessarily to divergencies.
In Appendix \ref{sec:Regularization-of-} we show exemplarily the
proper calculation for $\left\langle \vec{N}\vec{m}\left|V_{f^{+}\, b\, f^{-}}\right|\vec{N}\vec{m}'\right\rangle $.
We here give the regularized result for $\vec{m}=\vec{m}'$, since
it is of special importance for the discussion of the ground state
spectra away from half-filling, \begin{multline}
\left\langle \vec{N}\vec{m}\left|V_{f^{+}\, b\, f^{-}}\right|\vec{N}\vec{m}\right\rangle =u^{+}\sum_{r}\min(N_{r\uparrow},N_{r\downarrow})\\
+\frac{1}{4L}u^{+}\sum_{\{[r]_{f^{+}},[F]_{b},[\sigma]_{f^{-}}\}}\int dx\tilde{K}_{\vec{N}[r][F]}(x)\\
\times\left(\prod_{j\delta q}F(\lambda_{[r][F][\sigma]}^{j\delta q}(x),m_{j\delta q},m_{j\delta q})-1\right).\label{eq:VBSii_mm}\end{multline}

\section{The SWNT spectrum\label{sec:The-SWNT-spectrumresults}}

In Section \ref{sub:Diagonalizing} we wave diagonalized $H_{0}+V_{\rho\rho}$
and in Section \ref{sub:The-matrix-elements} we have determined the
matrix elements of $V_{\mathrm{n\rho\rho}}$ in the eigenbasis of
$H_{0}+V_{\rho\rho}$. Away from half-filling the magnitude of $V_{\mathrm{n}\rho\rho}$
is only small compared to $H_{0}+V_{\rho\rho}$ and therefore we can
easily analyze the effect of the non-density-density interaction $V_{\mathrm{n}\rho\rho}$
on the SWNT spectrum by representing the total Hamiltonian $H_{0}+V_{\rho\rho}+V_{\mathrm{n}\rho\rho}$
in a truncated eigenbasis of $H_{0}+V_{\rho\rho}$.

\subsection{The low energy spectrum away from half-filling\label{sub:The-low-energy}}

We start with the examination of the ground and low energy states.
As basis we use the lowest lying eigenstates of $H_{0}+V_{\rho\rho}$
without bosonic excitations with a given number of electrons $N_{c}$.%
\begin{figure}
\begin{center}\includegraphics[%
  width=0.42\columnwidth,
  angle=-90]{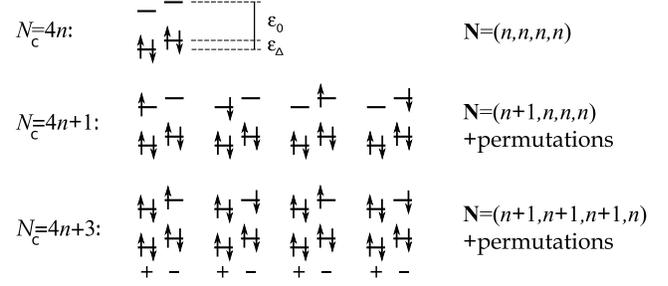}\end{center}

\caption{\label{cap:The-lowest-lying} The lowest lying eigenstates of $H_{0}+V_{\rho\rho}$
without bosonic excitations for the charge states $N_{c}=4n$, $N_{c}=4n+1$
and $N_{c}=4n+3$. On the right side the fermionic configurations
are given. We use the convention $\vec{N}=(N_{+\uparrow},N_{+\downarrow},N_{-\uparrow},N_{-\downarrow})$.}
\end{figure}

\subsubsection{$N_{c}=4n,$ $N_{c}=4n+1,$ $N_{c}=4n+3$}

First we consider the charge states $N_{c}=4n,$ $N_{c}=4n+1$ and
$N_{c}=4n+3$. In that case the lowest lying eigenstates of $H_{0}+V_{\rho\rho}$,
shown in Fig. \ref{cap:The-lowest-lying}, which are of the form $\left|\vec{N},0\right\rangle $
and therefore uniquely characterized by $\vec{N}$, do not mix via
$V_{\mathrm{n}\rho\rho}.$ That means that the only correction from
$V_{\mathrm{n}\rho\rho}$ to $H_{0}+V_{\rho\rho}$ stems from the
$\vec{N}$ conserving process $V_{f^{+}\, b\, f^{-}}$. For states
without bosonic excitations, equation (\ref{eq:VBSii_mm}) yields,
because of $F(\lambda,0,0)=1$, \begin{multline}
\left\langle \vec{N},0\left|V_{\mathrm{n}\rho\rho}\right|\vec{N},0\right\rangle =\left\langle \vec{N},0\left|V_{f^{+}\, b\, f^{-}}\right|\vec{N},0\right\rangle =\\
u^{+}\sum_{r}\min(N_{r\uparrow},N_{r\downarrow}).\label{eq:ME_NO}\end{multline}
Hence here $V_{\mathrm{n}\rho\rho}$ yields an energy penalty for
occupying the same branch $r$. This effect has already been found
in the meanfield theory of Oreg et al. \cite{Oreg2000}. The parameter
$\delta U$ there corresponds to our constant $u^{+}$. The energies
of the lowest lying states for $N_{c}=4n,$ $N_{c}=4n+1$ and $N_{c}=4n+3$
only depend on $\vec{N}.$ In detail we find with (\ref{eq:H0Vrr_diag})
and (\ref{eq:ME_NO}),\begin{multline}
E_{\vec{N}}=\frac{1}{2}E_{c}\mathcal{N}_{c}^{2}+u^{+}\sum_{r}\min(N_{r\uparrow},N_{r\downarrow})\\
+\frac{1}{2}\sum_{r\sigma}\mathcal{N}_{r\sigma}\left[-\frac{J}{2}\mathcal{N}_{-r\sigma}+\left(\varepsilon_{0}-u^{+}\right)\mathcal{N}_{r\sigma}+r\varepsilon_{\Delta}\right].\label{eq:E_N}\end{multline}
From (\ref{eq:E_N}) it follows that for the states depicted in Fig.
\ref{cap:The-lowest-lying} the interaction dependent part of $E_{\vec{N}}$
is the same for all fermionic configurations $\vec{N}$ corresponding to a given
charge state $N_{c}$. Hence the interaction leads merely to a common
shift of the lowest lying energy levels for fixed $N_{c}$.

\subsubsection{$N_{c}=4n+2$}

\begin{figure}
\begin{center}\includegraphics[%
  width=1.0\columnwidth]{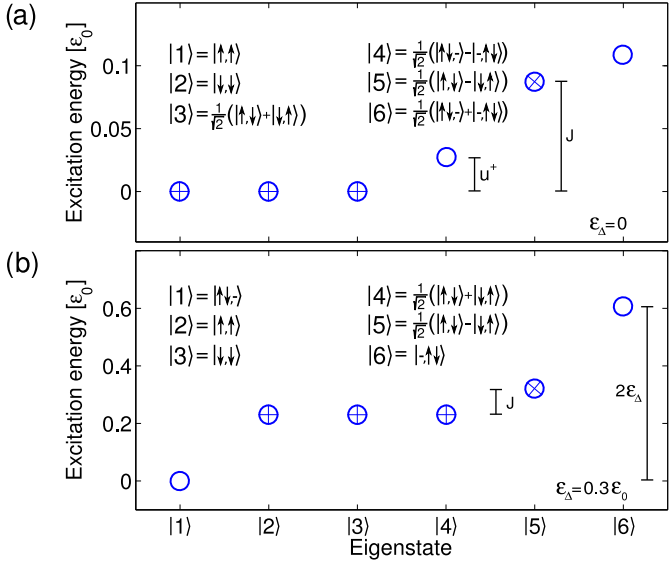}\end{center}

\caption{\label{cap:groundstatespectrum}Low energy spectrum of a $(6,6)$
SWNT for the charge state $N_{c}=4n+2$. (a) In the case $\varepsilon_{\Delta}=0$
the ground state is formed by the spin $1$ triplet $(\rightarrow\oplus)$
and the states $\left|\uparrow\downarrow,-\right\rangle $ and $\left|-,\uparrow\downarrow\right\rangle $
mix ($\rightarrow$\textit{\footnotesize $\bigcirc$} states). (b)
For $\varepsilon_{\Delta}\gg J/2$ the ground state is given by the
spin $0$ state $\left|\uparrow\downarrow,-\right\rangle $. }

The spin $0$ singlet state $1/\sqrt{2}\left(\left|\uparrow,\downarrow\right\rangle -\left|\downarrow,\uparrow\right\rangle \right)$
is indicated by $\otimes$. The coupling parameters are $J=0.09\varepsilon_{0}$
and $u^{+}\approx0.03\varepsilon_{0}$.
\end{figure}

Of special interest is the ground state structure of the $N_{c}=4n+2$
charge state, since here the lowest lying six eigenstates of $H_{0}+V_{\rho\rho}$
without bosonic excitations, denoted $\left|\vec{N},0\right\rangle $
with $\vec{N}=(n+1,n+1,n,n)+\textrm{permutations,}$ mix via $V_{\mathrm{n}\rho\rho}$,
leading to a $S=1$ triplet state and to three nondegenerate states
with spin $0$. For $\varepsilon_{\Delta}\approx0$ (the meaning of
$\approx0$ will become clear in the following) the triplet is the
ground state. In the following we are going to denote $\left|(n+1,n+1,n,n),0\right\rangle $
by $\left|\uparrow\downarrow,-\right\rangle $, $\left|(n+1,n,n,n+1),0\right\rangle $
by $\left|\uparrow,\downarrow\right\rangle $ and analogously for
the remaining four states. Ignoring interactions, the six considered
states are degenerate for $\varepsilon_{\Delta}=0$. As we can conclude
from (\ref{eq:H0Vrr_diag}) the degeneracy of the six considered states
is already lifted if including only the density-density interaction
$V_{\rho\rho}$, since then the energy of the spin $1$ states $\left|\uparrow,\uparrow\right\rangle $
and $\left|\downarrow,\downarrow\right\rangle $ is lowered by \begin{equation}
J/2:=u^{\Delta}_{f}+u^{\Delta}_{b}\label{eq:Def_J}\end{equation}
 relatively to the other ground states. Let us now consider the effects
of $V_{\mathrm{n}\rho\rho}$. The diagonal matrix elements $\left\langle \vec{N},0\left|V_{\mathrm{n}\rho\rho}\right|\vec{N},0\right\rangle $
are again determined by equation (\ref{eq:ME_NO}), leading to a relative
energy penalty for the states $\left|\uparrow\downarrow,-\right\rangle $
and $\left|-,\uparrow\downarrow\right\rangle $. Mixing occurs between
the states $\left|\uparrow,\downarrow\right\rangle $ and $\left|\downarrow,\uparrow\right\rangle $
via $V_{b\, f^{+}\, f^{-}}$and $V_{b\, b\, f^{-}}$ and between $\left|\uparrow\downarrow,-\right\rangle $
and $\left|-,\uparrow\downarrow\right\rangle $ via $V_{u\, f^{-}\, f^{-}}$
and $V_{u\, b\, f^{-}}$. With equation (\ref{eq:ME_V_nrr_expl})
we find \[
\left\langle \uparrow,\downarrow\left|V_{\mathrm{n}\rho\rho}\right|\downarrow,\uparrow\right\rangle =-\frac{J}{2}=-\left\langle \uparrow\downarrow,-\left|V_{\mathrm{n}\rho\rho}\right|-,\uparrow\downarrow\right\rangle .\]
In total, the SWNT Hamiltonian $H=H_{0}+V_{\rho\rho}+V_{\mathrm{n}\rho\rho}$
restricted to the basis spanned by the six states $\left|\uparrow,\uparrow\right\rangle $,
$\left|\downarrow,\downarrow\right\rangle $, $\left|\uparrow,\downarrow\right\rangle $,
$\left|\downarrow,\uparrow\right\rangle $, $\left|\uparrow\downarrow,-\right\rangle $
and $\left|-,\uparrow\downarrow\right\rangle $ is represented by
the matrix, \begin{multline}
H=E_{0,4n+2}+\\
\left(\begin{array}{ccccccc}
-\frac{J}{2} &  &  &  &  &  & 0\\
 &  & -\frac{J}{2}\\
 &  &  & 0 & -\frac{J}{2}\\
 &  &  & -\frac{J}{2} & 0\\
 &  &  &  &  & u^{+}-\varepsilon_{\Delta} & \frac{J}{2}\\
0 &  &  &  &  & \frac{J}{2} & u^{+}+\varepsilon_{\Delta}\end{array}\right),\label{eq:H_4mp2}\end{multline}
where $E_{0,4n+2}=\frac{1}{2}E_{c}N_{c}^{2}+(2n^{2}+2n+1)\left(\varepsilon_{0}-u^{+}\right)-J(n^{2}+n)+2u^{+}n$.
Diagonalizing the matrix in (\ref{eq:H_4mp2}), we find that its eigenstates
are given by the spin $1$ triplet \[
\left|\uparrow,\uparrow\right\rangle ,\left|\uparrow,\uparrow\right\rangle ,1/\sqrt{2}\left(\left|\uparrow,\downarrow\right\rangle +\left|\downarrow,\uparrow\right\rangle \right),\]
the spin $0$ singlet \[
1/\sqrt{2}\left(\left|\uparrow,\downarrow\right\rangle -\left|\downarrow,\uparrow\right\rangle \right)\]
 and the two states \[
\frac{1}{\sqrt{c_{1/2}^{2}+1}}\left(c_{1/2}\left|\uparrow\downarrow,-\right\rangle \pm\left|-,\uparrow\downarrow\right\rangle \right),\]
 where the coefficients $c_{1/2}$ are given by \[
c_{1/2}=\frac{\sqrt{\varepsilon_{\Delta}^{2}+(J/2)^{2}}\mp\varepsilon_{\Delta}}{J/2}.\]
Relatively to $E_{0,4n+2}$, the corresponding eigenenergies are $-J/2$
for the triplet states, $J/2$ for the singlet state and $u^{+}\pm\sqrt{\varepsilon_{\Delta}^{2}+(J/2)^{2}}$
for the remaining two states. Thus under the condition \[
J/2>\sqrt{\varepsilon_{\Delta}^{2}+(J/2)^{2}}-u^{+},\]
i.e., for a small band mismatch $\varepsilon_{\Delta}\lesssim J/2$
the ground state is degenerate and formed by the spin $1$ triplet,
otherwise by $\frac{1}{\sqrt{c_{2}^{2}+1}}\left(c_{2}\left|\uparrow\downarrow,-\right\rangle +\left|-,\uparrow\downarrow\right\rangle \right)$.
The ground state spectra for the two cases $\varepsilon_{\Delta}=0$
and $\varepsilon_{\Delta}\gg J/2$ are shown in Fig. \ref{cap:groundstatespectrum}
for a (6,6) armchair SWNT (corresponding to a diameter of $0.8$ nm).%
 Assuming a dielectric constant of $\varepsilon=1.4$ \cite{Egger1997},
the calculation of the coupling parameters according to Appendix \ref{sec:Modeling-the-interaction}
yields values of $J=2(u^{\Delta}_{f}+u^{\Delta}_{b})=0.09\varepsilon_{0}$
and $u^{+}\approx0.03\varepsilon_{0}$ which agree well with the experiments
\cite{Liang2002,Moriyama2005}, where nanotubes with $\varepsilon_{\Delta}\gg J/2$
were considered. To our knowledge, experiments in the regime $\varepsilon_{\Delta}\lesssim J/2$
demonstrating exchange effects have not been carried out so far,
such that a validation of our predictions for this case, namely the
existence of the ground state spin $1$ triplet and the mixing of
the states $\left|\uparrow\downarrow,-\right\rangle $ and $\left|-,\uparrow\downarrow\right\rangle $
is still missing. The latter effect could be of relevance for the understanding
of the so called singlet-triplet Kondo effect \cite{EtoNazarov2000}
in SWNTs.

It should be stressed that all exchange effects, leading amongst others
to the spin $1$ triplet as ground state, result from $S_{r}\neq f$
interaction processes. In the work of Mattis and Lieb \cite{Lieb-Mattis1962}
however, there is no such additional pseudo spin degree of freedom.
Hence we suspect that this is the reason why their theorem can not
be applied in our situation.

\subsection{Excitation spectra away from half-filling\label{sub:Excitation-spectra-away}}

\begin{figure}
\begin{center}\includegraphics[%
  width=1.0\columnwidth]{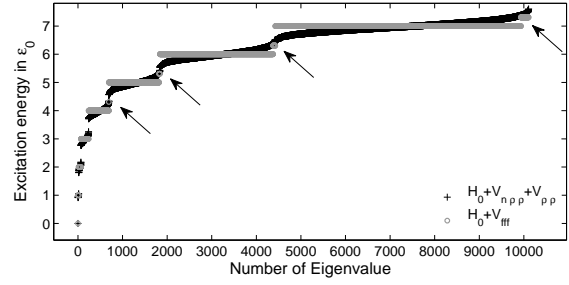}\end{center}

\caption{\label{H_st_vs_H_full}The excitation spectrum for a (6,6) SWNT occupied
by $N_{c}=4n$ electrons. In grey we show the spectrum as obtained
by diagonalizing the Hamiltonian of the standard theory $H_{st}=H_{0}+V_{f\, f\, f}$
and in black for the full Hamiltonian $H=H_{0}+V_{\rho\rho}+V_{\mathrm{n}\rho\rho}$.
A band mismatch $\varepsilon_{\Delta}=0$ is assumed. The energy of
the lowest $c+$ excitation is $4.3\varepsilon_{0}$. All other interaction
parameters are as in Fig. \ref{cap:groundstatespectrum}. Arrows indicate
eigenenergies of the {}``standard'' Hamiltonian $H_{st}=H_{0}+V_{f\, f\, f}$
involving excitations of the $c+$ mode.}
\end{figure}

\begin{figure}
\begin{center}\includegraphics[%
  width=1.0\columnwidth]{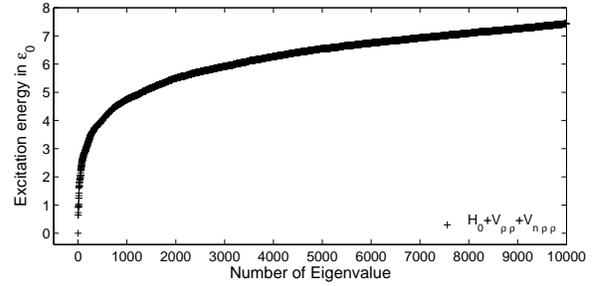}\end{center}

\caption{\label{cap:eps_Delta}The excitation spectrum for a (6,6) SWNT obtained
by diagonalizing the full Hamiltonian $H=H_{0}+V_{\rho\rho}+V_{\mathrm{n}\rho\rho}$
for $N_{c}=4m$ and $\varepsilon_{\Delta}=0.3\varepsilon_{0}$. The
spectrum becomes quasicontinuous at relatively small energies. Shown
are the lowest $10000$ eigenenergies.}
\end{figure}

\begin{figure}
\begin{center}\includegraphics[%
  width=1.0\columnwidth]{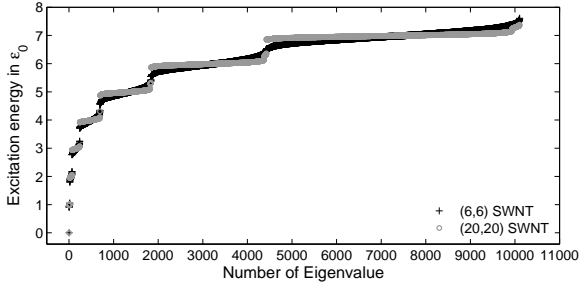}\end{center}

\caption{\label{cap:66vs2020}The excitation spectrum for a (6,6) SWNT (black)
compared to the spectrum of a (20,20) SWNT (grey). The effects of
the non forward scattering processes are by far less pronounced in
the latter case. $N_{c}=4m$ and $\varepsilon_{\Delta}=0$.}
\end{figure}

\begin{figure}
\begin{center}\includegraphics[%
  width=1.0\columnwidth]{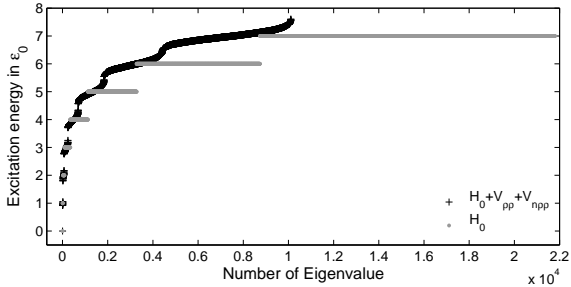}\end{center}

\caption{\label{cap:nointvsfullint} The excitation spectrum for a (6,6) SWNT
obtained by diagonalizing the Hamiltonian of the noninteracting system
$H_{0}$ (grey) and the full Hamiltonian $H=H_{0}+V_{\rho\rho}+V_{\mathrm{n}\rho\rho}$
(black).}
\end{figure}

Until now our discussion of the energy spectra was based on states
$\left|\vec{N},0\right\rangle $ without bosonic excitations and
so far the effect of $V_{\mathrm{n}\rho\rho}$ on the spectrum
could have even been treated without using bosonization. But for the determination
of the excitation spectrum of $H$ we do need the general expression
for the matrix elements of $V_{\mathrm{n}\rho\rho}$ between the eigenstates
of $H_{0}+V_{\rho\rho}$ as given by (\ref{eq:ME_V_nrr_expl}). For
the actual calculation we truncate the eigenbasis of $H_{0}+V_{\rho\rho}$
for a fixed charge state $N_{c}$ at a certain excitation energy and
represent $H$ in this shortened basis. After the diagonalization
we find to a good approximation the correct eigenstates and eigenenergies
of $H.$ For the results shown in Figs. \ref{H_st_vs_H_full} to \ref{cap:nointvsfullint}
we have checked that convergence has been reached, i.e., the extention
of the considered basis states does not lead to a significant change
of the spectrum. 

Exemplarily we present the results for the charge state $N=4n.$ Similar
excitation spectra are found for the other charge states. In Fig.
\ref{H_st_vs_H_full} we show for comparison and in order to demonstrate
the effect of the non forward scattering processes the findings for
the {}``standard'' theory, i.e., the spectrum of $H_{st}=H_{0}+V_{f\, f\, f}$
as well as the spectrum of the full Hamiltonian $H=H_{0}+V_{\rho\rho}+V_{\mathrm{n}\rho\rho}$
for a $(6,6)$ armchair nanotube. Thereby a nonvanishing band mismatch
$\varepsilon_{\Delta}=0$ is assumed. Striking is the partial breaking
of the huge degeneracies of the {}``standard'' spectrum. Note also
the lifting of the spin-charge separation when including the non forward
scattering processes. To illustrate this point we have indicated eigenenergies
of $H_{st}$ including $c+$ excitations by arrows in Fig. \ref{H_st_vs_H_full}.

At higher energies a quasi continuum forms in the case of the full
Hamiltonian $H$, a feature becoming especially apparent for a finite
band mismatch. In Fig. \ref{cap:eps_Delta} the spectra of the full
Hamiltonian $H$ is shown for $\varepsilon_{\Delta}=0.3\varepsilon_{0}$.

As we have already discussed, the importance of non forward scattering
terms should decrease with increasing tube diameter. And indeed the
excitation spectrum of the full Hamiltonian for a $(20,20)$ SWNT
resembles much more the result of the {}``standard'' theory than
it is the case for a $(6,6)$ SWNT as it can be seen from Fig. \ref{cap:66vs2020}.%

It is also interesting to regard the effect of the total interaction
$V_{\rho\rho}+V_{\mathrm{n}\rho\rho}$ on the nanotube spectrum. For
this purpose, in Fig. \ref{cap:nointvsfullint} the spectrum of $H_{0}$
describing the noninteracting system is compared to the spectrum of
the full Hamiltonian $H,$ again for a $(6,6)$ SWNT with vanishing
band mismatch. Of special significance is the strong reduction of
the number of eigenstates below a certain energy if the interaction
is {}``switched on''. This can be mainly traced back to $V_{fff}$
which leads to the formation of the bosonic $c+$ excitations with
considerably enlarged energies. Concerning the transport properties
of SWNTs the reduction of relevant states plays an important role
for the occurrence of the power law dependence of various transport
quantities in the case of infinitely long tubes but also for the appearance
of negative differential conductance in highly asymmetric SWNT quantum
dots as described in \cite{Mayrhofer2006}.%

\subsection{Comparison to the mean field results}

We shortly want to compare the results of the meanfield theory by
Oreg et al. \cite{Oreg2000} and our approach. Concerning the groundstate
structure, differences between the two works arise for the $N_{c}=4m+2$
charge state. In this situation the meanfield Hamiltonian can essentially
be recovered by setting all off-diagonal elements in (\ref{eq:H_4mp2})
to zero. Therefore in \cite{Oreg2000} the degenerate triplet state
can not be predicted but twofold degeneracies of the states $\left|\uparrow,\uparrow\right\rangle $,
$\left|\downarrow,\downarrow\right\rangle $ and of $\left|\uparrow,\downarrow\right\rangle $,
$\left|\downarrow,\uparrow\right\rangle $ respectively are found.
Moreover contrary to our theory in \cite{Oreg2000} no mixing of the
states $\left|\uparrow\downarrow,-\right\rangle $ and $\left|-,\uparrow\downarrow\right\rangle $
can occur for $\varepsilon_{\Delta}\lesssim J/2,$ an important point
regarding the singlet-triplet Kondo effect \cite{EtoNazarov2000}. 

Moreover also the excitation spectrum shows considerable differences
in both approaches, since the meanfield approach misses the formation
of the collective electronic excitations as the $c+$ mode, with its
dispersion relation strongly renormalized by the forward scattering
part of the Coulomb interaction.

\subsection{Near half-filling}

As we have already seen in Section \ref{sub:The-matrix-elements},
at half-filling non-density-density interaction processes become relevant
which yield tremendously large matrix elements in the eigenbasis of
$H_{0}+V_{\rho\rho}$, as a consequence of the function $A_{u}(x)$
shown in Fig. \ref{cap:A_hf}. Therefore our truncation scheme for
diagonalizing the total Hamiltonian $V_{n\rho\rho}$ does not give
reliable results at half-filling. Investigation of the half-filling
case is beyond the scope of this work.

\section{Conclusions}

In summary, we have derived the low energy Hamiltonian for metallic
finite size SWNTs including all relevant interaction terms, especially
the short ranged processes whose coupling strength scales inversely
proportional to the SWNT size. The Hamiltonian of the noninteracting
system, $H_{0}$, together with the density-density part of the interaction,
$V_{\rho\rho}$, could be diagonalized by bosonization and Bogoliubov
transformation. Considering only the situation away from half-filling,
we obtained the spectrum of the total SWNT Hamiltonian by exploiting
the small magnitude of the non-density-density contribution $V_{\mathrm{n}\rho\rho}$
to the interaction: we have calculated the matrix elements of $V_{\mathrm{n}\rho\rho}$
in a truncated eigenbasis of $H_{0}+V_{\rho\rho}$ and diagonalized
the resulting matrix to obtain the SWNT spectrum and the corresponding
eigenstates. 

Of special interest, concerning the ground state spectra, is the formation
of a spin $1$ triplet for the charge state $N_{c}=4m+2$, whose existence
has clearly been proven in the experiments of Moriyama et al. \cite{Moriyama2005}.
In the case of a band mismatch $\varepsilon_{\Delta}$ that is small
compared to the exchange energy $J,$ the spin $1$ triplet is the
ground state of the system. This finding is interesting since according
to a theorem by Lieb and Mattis \cite{Lieb-Mattis1962}, only ground
states with spin $0$ or $1/2$ are allowed for a 1D Hubbard model
with next-neighbour hopping and no orbital degeneracies. Since our
SWNT Hamiltonian includes an orbital degree of freedom we conclude
that scattering processes with respect to this degree of freedom are
the reason for the finding of a spin $1$ ground state. Additionally
we predict for $\varepsilon_{\Delta}\lesssim J/2$, the mixing of
the states $\left|\uparrow\downarrow,-\right\rangle $ and $\left|-,\uparrow\downarrow\right\rangle $
with an accompanying energy splitting. The degree of mixing between
$\left|\uparrow\downarrow,-\right\rangle $ and $\left|-,\uparrow\downarrow\right\rangle $
is of importance for the singlet-triplet Kondo effect, as discussed
in \cite{EtoNazarov2000}. An experimental confirmation of our findings
in the case $\varepsilon_{\Delta}\lesssim J/2$ is still missing,
but well within reach. 

With regard to the excitation spectrum, the different bosonic modes
are mixed by the non-density-density interaction processes $V_{\mathrm{n\rho\rho}}$.
Therefore the spin-charge separation is lifted. Moreover we find that
the huge degeneracies which are obtained by the {}``standard'' theory
that retains only forward scattering processes are partially broken.
This leads to a more and more continuous energy spectrum for higher
energies.

\appendix

\section{Modelling the interaction potential\label{sec:Modeling-the-interaction}}

In this Appendix we show how we determine the values of the effective
1D potentials $U_{[F]}^{intra/inter}$ and of the coupling constants
$u^{S_{r}\, S_{F}}$. We start with equation (\ref{eq:Uintra_INTER})
from section \ref{sec:The-interaction-Hamiltonian}, \begin{multline}
U_{[F]}^{intra/inter}(x,x')=L^{2}\int\int d^{2}r_{\perp}d^{2}r'_{\perp}\\
\times\varphi_{pF_{1}}^{*}(\vec{r})\varphi_{\pm pF_{2}}^{*}(\vec{r}')\varphi_{\pm pF_{3}}(\vec{r}')\varphi_{pF_{4}}(\vec{r})U(\vec{r}-\vec{r}').\label{eq:Uintra_INTER_App}\end{multline}
Using equation (\ref{eq:phi_pF_chi}) in order to reexpress the Bloch
waves $\varphi_{pF}(\vec{r})$ in terms of $p_{z}$ orbitals, we obtain,\begin{multline}
U_{[F]}^{intra/inter}(x,x')=\frac{L^{2}}{N_{L}^{2}}\int\int d^{2}r_{\perp}d^{2}r'_{\perp}\\
\times U(\vec{r}-\vec{r}')\sum_{\vec{R},\vec{R}'}e^{-i\left(F_{1}-F_{4}\right)R_{x}-i\left(F_{2}-F_{3}\right)R'_{x}}\\
\times\left|\chi(\vec{r}-\vec{R}-\vec{\tau}_{p})\right|^{2}\left|\chi(\vec{r}'-\vec{R}'-\vec{\tau}_{\pm p})\right|^{2}.\label{eq:Uinterintra_pz_App}\end{multline}
Instead of a fourfold sum over the lattice sites only the double sum
$\sum_{\vec{R},\vec{R}'}$ remains, since the overlap of different
$p_{z}$ orbitals can be neglected. To proceed we use once more that
the spatial extention of the $p_{z}$ orbitals is small compared to
all other appearing length scales and therefore replace $\left|\chi(\vec{r}-\vec{R}-\vec{\tau}_{p})\right|^{2}$
by the delta function $\delta(\vec{r}-\vec{R}-\vec{\tau}_{p}).$ In
order to take into account the error induced thereby at small distances
$x\approx x'$, we replace the Coulomb potential by the Ohno potential
introduced by equation (\ref{eq:Ohno}). It interpolates between $U_{0},$
the interaction energy between two $p_{z}$ electrons in the same
orbital and $\frac{e^{2}}{4\pi\epsilon_{0}\epsilon\left|\vec{r}-\vec{r}\,'\right|}$
for $\left|\vec{r}-\vec{r}\,'\right|\gg0$. Performing the integration
in (\ref{eq:Uinterintra_pz_App}), we obtain,\begin{multline}
U_{[F]}^{intra/inter}(x,x')=\frac{L^{2}}{N_{L}^{2}}\sum_{\vec{R},\vec{R}'}\delta(x-R_{x})\delta(x'-R'_{x})\\
\times e^{-i\left(F_{1}-F_{4}\right)R_{x}-i\left(F_{2}-F_{3}\right)R'_{x}}U(\vec{R}-\vec{R}'+\vec{\tau}_{p}-\vec{\tau}_{\pm p}).\label{eq:Uintrainter_discrete}\end{multline}
 Now we can easily calculate the values of the coupling constants
$u^{S_{r}\, S_{F}}$ for the local interactions, given by (\ref{eq:couplingparameters}),\[
u^{S_{r}\, S_{F}}=1/(2L^{2})\int\int dx\, dx'U_{[r]_{S_{r}}[F]_{S_{F}}}(x,x').\]
 Using (\ref{eq:Uintrainter_discrete}) together with equation (\ref{eq:U_rF_Uint}),\begin{multline}
U_{[r][F]}(x,x')=\frac{1}{4}\left[U_{[F]}^{intra}(x,x')(1+r_{1}r_{2}r_{3}r_{4})\right.\\
+\left.U_{[F]}^{inter}(x,x')(r_{2}r_{3}+r_{1}r_{4})\right],\label{eq:U_rF_Uint_app}\end{multline}
 we arrive at\begin{multline}
u^{f\, b}=:u^{+}=\frac{1}{4N_{L}^{2}}\sum_{\vec{R},\vec{R}'}e^{-i2K_{0}(R_{x}-R'_{x})}\\
\times\left[U(\vec{R}-\vec{R}')+U(\vec{R}-\vec{R}'+\vec{\tau}_{p}-\vec{\tau}_{-p})\right],\label{eq:u+}\end{multline}
\begin{multline}
u^{b/u\, f}=:u^{\Delta}_{f}=\\
\frac{1}{4N_{L}^{2}}\sum_{\vec{R},\vec{R}'}\left[U(\vec{R}-\vec{R}')-U(\vec{R}-\vec{R}'+\vec{\tau}_{p}-\vec{\tau}_{-p})\right]\label{eq:udelta_f}\end{multline}
and \begin{multline}
u^{b/u\, f}=:u^{\Delta}_{b}=\frac{1}{4N_{L}^{2}}\sum_{\vec{R},\vec{R}'}e^{-i2K_{0}(R_{x}-R'_{x})}\\
\times\left[U(\vec{R}-\vec{R}')-U(\vec{R}-\vec{R}'+\vec{\tau}_{p}-\vec{\tau}_{-p})\right].\label{eq:udelta_b}\end{multline}
 Since in the summations in (\ref{eq:u+}), (\ref{eq:udelta_f}) and
(\ref{eq:udelta_b}) only terms with $\vec{R}\approx\vec{R}'$ contribute,
the number of relevant summands scales like the number of lattice
sites $N_{L}$. Due to the prefactor $1/N_{L}^{2}$, $u^{+}$ and
$u^{\Delta}_{f/b}$ in total scale like $1/N_{L}$. Numerical evaluation
of the previous three equations leads to the values given in table
\ref{cap:us}.

\section{\label{sec:CalcMmm}Calculation of the matrix elements $M_{[r][F][\sigma]}(\vec{N},\vec{m},\vec{N}',\vec{m}',x)$}

Using the bosonization identity (\ref{eq:bosident}),\[
\psi_{r\sigma F}(x)=\eta_{r\sigma}K_{r\sigma F}(x)e^{i\phi_{r\sigma F}^{\dagger}(x)}e^{i\phi_{r\sigma F}(x)},\]
 we can separate $M_{[r][F][\sigma]}(\vec{N},\vec{m},\vec{N}',\vec{m}',x)$
from equation (\ref{eq:DefMNmNmx}) into a bosonic and a fermionic
part,\begin{multline*}
M_{[r][F][\sigma]}(\vec{N},\vec{m},\vec{N}',\vec{m}',x)=\\
M_{[r][F][\sigma]}(\vec{N},\vec{N}',x)M_{[r][F][\sigma]}(\vec{m},\vec{m}',x),\end{multline*}
where \begin{multline}
M_{[l]}(\vec{N},\vec{N}',x)=\\
\left\langle \vec{N}\right|K_{l_{1}}^{\dagger}(x)\eta_{l_{1}}^{\dagger}K_{l_{2}}^{\dagger}(x)\eta_{l_{2}}^{\dagger}K_{l_{3}}(x)\eta_{l_{3}}K_{l_{4}}(x)\eta_{l_{4}}\left|\vec{N}'\right\rangle \end{multline}
and \begin{multline}
M_{[l]}(\vec{m},\vec{m}',x)=\left\langle \vec{m}\right|e^{-i\phi_{l_{1}}^{\dagger}(x)}e^{-i\phi_{l_{1}}(x)}e^{-i\phi_{l_{2}}^{\dagger}(x)}e^{-i\phi_{l_{2}}(x)}\\
e^{i\phi_{l_{3}}^{\dagger}(x)}e^{i\phi_{l_{3}}(x)}e^{i\phi_{l_{4}}^{\dagger}(x)}e^{i\phi_{l_{4}}(x)}\left|\vec{m}'\right\rangle .\label{eq:Mmm_def}\end{multline}
Improving readability, we have summarized the indices $rF\sigma$
by a single index $l$.

\subsection{The Fermionic part of $M_{[r][F][\sigma]}(\vec{N},\vec{m},\vec{N}',\vec{m}',x)$}

First we consider the contribution $M_{[l]}(\vec{N},\vec{N}',x)$
depending on the fermionic configurations $\vec{N}$ and $\vec{N}'$.
Using relation (\ref{eq:Def_Klein_factors}) for the Klein factors
$\eta_{r\sigma}$ and the definition of the phase factor $K_{r\sigma F}(x)$,
equation (\ref{eq:K_rsF}), we obtain \begin{multline*}
M_{[r][F][\sigma]}(\vec{N},\vec{N}',x)=\\
\frac{1}{(2L)^{2}}\delta_{\vec{N},\vec{N}'+\vec{E}_{[r][\sigma]}}T_{\vec{N}\vec{N}'[r][\sigma]}Q_{\vec{N}\vec{N}'[r][F]}(x),\end{multline*}
where $\vec{E}_{[r][\sigma]}:=\vec{e}_{r_{1}\sigma}+\vec{e}_{r_{2}\sigma'}-\vec{e}_{r_{3}\sigma'}-\vec{e}_{r_{4}\sigma}$.
Furthermore $T_{\vec{N}\vec{N}'[r][\sigma]}$ is given by \begin{multline}
T_{\vec{N}\vec{N}'[r][\sigma]}=(-1)^{\sum_{j_{4}=1}^{(r_{4}\sigma_{4})-1}(\vec{N}')_{j_{4}}+\sum_{j_{3}=1}^{(r_{3}\sigma_{3})-1}(\vec{N}'-\vec{e}_{r_{4}\sigma_{4}})_{j_{3}}}\\
\times(-1)^{\sum_{j_{2}=1}^{(r_{2}\sigma_{2})-1}(\vec{N}-\vec{e}_{r_{1}\sigma_{1}})_{j_{2}}+\sum_{j_{1}=1}^{(r_{1}\sigma_{1})-1}(\vec{N})_{j_{1}}}.\label{eq:TNNp}\end{multline}
Here we use the convention $j=+\uparrow,+\downarrow,-\uparrow,-\downarrow\,=1,2,3,4$.
It turns out that $T_{\vec{N}\vec{N}'[r][\sigma]}$ only depends on
the scattering types $S_{r}$ and $S_{\sigma}$. Explicitly with $T_{\vec{N}'S_{r}S_{\sigma}}:=T_{\vec{N}\vec{N}'[r]_{S_{r}}[\sigma]_{S_{\sigma}}}$,
\begin{equation}
T_{\vec{N}'uf^{-}}=-(-1)^{3N'_{R\uparrow}+2N'_{R\downarrow}+N'_{L\uparrow}},\label{eq:T_Nuf}\end{equation}
\begin{equation}
T_{\vec{N}'bf^{-}}=(-1)^{3N'_{R\uparrow}+2N'_{R\downarrow}+N'_{L\uparrow}}\label{eq:T_Nbf}\end{equation}
and $T_{\vec{N}'S_{r}S_{\sigma}}=1$ for all other $(S_{r},S_{\sigma})$.
Finally the function $Q_{\vec{N}\vec{N}'[r][F]}(x)$ yields a phase and is
given by \begin{multline}
Q_{\vec{N}\vec{N}'[r][F]}(x)=\\
\exp\left\{ i\frac{\pi}{L}\left[\mathrm{sgn}(r_{4}F_{4})(\vec{N}')_{l_{4}}+\mathrm{sgn}(r_{3}F_{3})(\vec{N}'-\hat{e}_{l_{4}})_{l_{3}}\right.\right.\\
\left.\left.-\mathrm{sgn}(r_{2}F_{2})(\vec{N}-\hat{e}_{l_{1}})_{l_{2}}-\mathrm{sgn}(r_{1}F_{1})(\vec{N})_{l_{1}}\right]x\right\} .\label{eq:Q_NrF}\end{multline}

\subsection{The bosonic part of $M_{[r][F][\sigma]}(\vec{N},\vec{m},\vec{N}',\vec{m}',x)$}

The calculation of the bosonic part $M_{[r][F][\sigma]}(\vec{m},\vec{m}',x)$
is based on expressing the fields $i\phi_{r\sigma F}(x)$ in equation
(\ref{eq:Mmm_def}) in terms of the bosonic operators $a_{j\delta q}$,
$a_{j\delta q}^{\dagger}$ and subsequent normal ordering, i.e., commuting
all annihilation operators $a_{j\delta q}$ to the right side and
all creation operators $a_{j\delta q}^{\dagger}$ to the left side.
In a first step we use the relation \[
e^{i\phi_{l}(x)}e^{i\phi_{l}^{\dagger}(x)}=e^{i\phi_{l}^{\dagger}(x)}e^{i\phi_{l}(x)}e^{[i\phi_{l}(x),i\phi_{l}^{\dagger}(x)]},\]
following from the Baker-Hausdorff formula \cite{Delft1998}, \[
e^{A}e^{B}=e^{A+B}e^{\frac{1}{2}[A,B]},\;[A,B]\in\mathbb{C},\]
to obtain from (\ref{eq:Mmm_def}),\begin{multline}
M_{[l]}(\vec{m},\vec{m}',x)=C_{[l]}(x)\\
\times\left\langle \vec{m}\left|e^{-i\tilde{\sum}_{n=1}^{4}\phi_{l_{n}}^{\dagger}(x)}e^{-i\tilde{\sum}_{n=1}^{4}\phi_{l_{n}}(x)}\right|\vec{m}'\right\rangle ,\label{eq:Mmminterm_1}\end{multline}
where $\tilde{\sum}_{l=1}^{4}\phi_{l_{n}}$ denotes the sum $\phi_{l_{1}}+\phi_{l_{2}}-\phi_{l_{3}}-\phi_{l_{4}}$
and \begin{multline*}
C_{[l]}(x)=e^{[i\phi_{l_{3}}(x),i\phi_{l_{4}}^{\dagger}(x)]}e^{[-i\phi_{l_{2}}(x),i\phi_{l_{3}}^{\dagger}(x)+i\phi_{l_{4}}^{\dagger}(x)]}\\
\times e^{[-i\phi_{l_{1}}(x),-i\phi_{l_{2}}^{\dagger}(x)+i\phi_{l_{3}}^{\dagger}(x)+i\phi_{l_{4}}^{\dagger}(x)]}.\end{multline*}
Applying the Baker-Hausdorff formula once more, we obtain \begin{multline*}
e^{-i\tilde{\sum}_{n=1}^{4}\phi_{l_{n}}^{\dagger}(x)}e^{-i\tilde{\sum}_{n=1}^{4}\phi_{l_{n}}(x)}=\\
e^{-i\tilde{\sum}_{n=1}^{4}\left(\phi_{l_{n}}(x)+\phi_{l_{n}}^{\dagger}(x)\right)}e^{\frac{1}{2}\left[i\tilde{\sum}_{n=1}^{4}\phi_{l_{n}}^{\dagger}(x),i\tilde{\sum}_{n'=1}^{4}\phi_{l_{n'}}(x)\right]}.\end{multline*}
Using the definition of the $\phi$-fields, equation (\ref{eq:phifield_b}),
together with the transformation between the operators $b_{\sigma q}$
and $a_{j\delta q}$, equation (\ref{eq:b_a}), we get\begin{multline*}
i\phi_{r\sigma F}(x)+i\phi_{r\sigma F}^{\dagger}(x)=\\
\sum_{j\delta q>0}\left(\lambda_{r\sigma F}^{j\delta q}(x)a_{j\delta q}-\lambda_{r\sigma F}^{*j\delta q}(x)a_{j\delta q}^{\dagger}\right).\end{multline*}
In terms of $\Lambda_{r\sigma}^{j\delta}$, $B_{j\delta q}$ and $D_{j\delta q}$,
cf. equations (\ref{eq:Lambdajd_rs}), (\ref{eq:SC_ntrl}) and (\ref{eq:SCcp}),
the coefficients $\lambda_{r\sigma F}^{j\delta q}(x)$ read\begin{equation}
\lambda_{r\sigma F}^{j\delta q}(x)=\frac{\Lambda_{r\sigma}^{j\delta}}{\sqrt{n_{q}}}\left(e^{i\mathrm{sgn}(rF)qx}B_{j\delta q}-e^{-i\mathrm{sgn}(rF)qx}D_{j\delta q}\right).\label{eq:lambda_einfach}\end{equation}
By defining \begin{equation}
\tilde{\lambda}_{[l]}^{j\delta q}(x):=-\tilde{\sum}_{n=1}^{4}\lambda_{l_{n}}^{j\delta q}(x)\label{eq:lambda_tilde}\end{equation}
and again using the Baker-Hausdorff formula, we arrive at \begin{multline*}
e^{-i\tilde{\sum}_{n=1}^{4}\left(\phi_{l_{n}}(x)+\phi_{l_{n}}^{\dagger}(x)\right)}=\\
e^{-\sum_{j\delta q>0}\tilde{\lambda}_{[l]}^{*j\delta q}(x)a_{j\delta q}^{\dagger}}e^{\sum_{j\delta q>0}\tilde{\lambda}_{[l]}^{j\delta q}(x)a_{j\delta q}}e^{-\frac{1}{2}\sum_{j\delta q>0}\left|\tilde{\lambda}_{[l]}^{j\delta q}(x)\right|^{2}},\end{multline*}
such that in total \begin{multline}
\left\langle \vec{m}\left|e^{-i\tilde{\sum}_{n=1}^{4}\phi_{l_{n}}^{\dagger}(x)}e^{-i\tilde{\sum}_{n=1}^{4}\phi_{l_{n}}(x)}\right|\vec{m}'\right\rangle =\\
A_{[l]}(x)\prod_{j\delta q}F(\tilde{\lambda}_{[l]}^{j\delta q}(x),m_{j\delta q},m'_{j\delta q}),\label{eq:Mmminterm_2}\end{multline}
where we have introduced \begin{multline}
A_{[l]}(x):=e^{\frac{1}{2}\left[i\tilde{\sum}_{n=1}^{4}\phi_{l_{n}}^{\dagger}(x),i\tilde{\sum}_{n'=1}^{4}\phi_{l_{n'}}(x)\right]}\\
\times e^{-\frac{1}{2}\sum_{j\delta q>0}\left|\tilde{\lambda}_{[l]}^{j\delta q}(x)\right|^{2}}.\label{eq:A_def}\end{multline}
The function $F(\lambda,m,m')=\left\langle m\left|e^{-\lambda a^{\dagger}}e^{\lambda a}\right|m'\right\rangle $
is given by \cite{Mayrhofer2006} \begin{multline}
F(\lambda,m,m')=\\
\left(\Theta(m'-m)\lambda^{m'-m}+\Theta(m-m')\left(-\lambda^{*}\right)^{m-m'}\right)\\
\times\sqrt{\frac{m_{min}!}{m_{max}!}}\sum_{i=0}^{m_{min}}\frac{\left(-\left|\lambda\right|^{2}\right)^{i}}{i!(i+m_{max}-m_{min})!}\frac{m_{max}!}{(m_{min}-i)!},\label{eq:Fvonlambda1}\end{multline}
where $m_{min/max}=\min/\max(m,m')$. Combining (\ref{eq:Mmminterm_1})
and (\ref{eq:Mmminterm_2}) we finally obtain\begin{multline*}
M_{[l]}(\vec{m},\vec{m}',x)=C_{[l]}(x)\\
\times A_{[l]}(x)\prod_{j\delta q}F(\tilde{\lambda}_{[l]}^{j\delta q}(x),m_{j\delta q},m'_{j\delta q}).\end{multline*}
Explicitly, equation (\ref{eq:A_def}) yields that $A_{[l]}(x)$ only
depends on the scattering type for the product $rF$. For $S_{rF}\neq u$
we find $A_{[l]_{S_{rF}}}=:A_{S_{rF}}\equiv1$ whereas $A_{u}$ is
strongly enhanced leading to an increased importance of non-density-density
interactions at half-filling. Due to its relevance we show the detailed
calculation of $A_{u}$ in the following.

\subsubsection{Evaluation of $A_{u}$}

As example we calculate $A_{[r]_{S_{r}}[F]_{S_{F}}[\sigma]_{S_{\sigma}}}$
with $(S_{r},S_{F},S_{\sigma})=(b,f^{-},f^{+})$, i.e., for $[r]=(r,-r,r,-r)$,
$[F]=(F,-F,-F,F)$ and $[\sigma]=(\sigma,\sigma,\sigma,\sigma)$.
It is easily checked that for this choice $S_{rF}=u$ holds. Before
starting with the actual calculation we first determine the coefficients
$\tilde{\lambda}_{[r][F][\sigma]}^{j\delta q}(x)$ for the considered
case. With equations (\ref{eq:lambda_einfach}) and (\ref{eq:lambda_tilde})
we find\begin{multline*}
\tilde{\lambda}_{[r]_{b}[F]_{f^{-}}[\sigma]_{f^{+}}}^{j\delta q}(x)=\\
-\frac{1}{\sqrt{n_{q}}}\tilde{\sum}_{n=1}^{4}\Lambda_{r_{n}\sigma_{n}}^{j\delta}\left(e^{i\mathrm{sgn}(r_{n}F_{n})qx}B_{j\delta q}-e^{-i\mathrm{sgn}(r_{n}F_{n})qx}D_{j\delta q}\right).\end{multline*}
 The values for $B_{j\delta q}$, $D_{j\delta q}$ and $\Lambda_{r\sigma}^{j\delta}$
are known from the Bogoliubov transformation, cf. equations (\ref{eq:Lambdajd_rs})
to (\ref{eq:SCcp}). For the different channels $j\delta$ this leads
to \[
\tilde{\lambda}_{[r]_{b}[F]_{f^{-}}[\sigma]_{f^{+}}}^{c+q}(x)=-\frac{2i\mathrm{sgn}(rF)}{\sqrt{n_{q}}}\sqrt{\frac{\varepsilon_{0q}}{\varepsilon_{c+q}}}\sin(qx),\]
\[
\tilde{\lambda}_{[r]_{b}[F]_{f^{-}}[\sigma]_{f^{+}}}^{c-q}(x)=0\]
 \[
\tilde{\lambda}_{[r]_{b}[F]_{f^{-}}[\sigma]_{f^{+}}}^{s+q}(x)=-\frac{2i\mathrm{sgn}(rF\sigma)}{\sqrt{n_{q}}}\sin(qx),\]
 \[
\tilde{\lambda}_{[r]_{b}[F]_{f^{-}}[\sigma]_{f^{+}}}^{s-q}(x)=0.\]
 Using (\ref{eq:A_def}) we get in this case,\begin{multline}
A_{[l]}(x):=e^{\frac{1}{2}\left[i\phi_{l_{1}}^{\dagger}(x)-i\phi_{l_{3}}^{\dagger}(x),i\phi_{l_{1}}(x)-i\phi_{l_{3}}(x)\right]}\\
\times e^{\frac{1}{2}\left[i\phi_{l_{2}}^{\dagger}(x)-i\phi_{l_{4}}^{\dagger}(x),i\phi_{l_{2}}(x)-i\phi_{l_{4}}(x)\right]}\\
\times e^{-\frac{1}{2}\sum_{q>0}\left(\left|\tilde{\lambda}_{[l]}^{c+q}(x)\right|^{2}+\left|\tilde{\lambda}_{[l]}^{s+q}(x)\right|^{2}\right)}.\label{eq:A_rFS}\end{multline}
Improving readability we have again replaced the indices $rF\sigma$
by a single index $l$. With (\ref{eq:phifield_b}) we obtain\begin{multline*}
\left[i\phi_{rF\sigma}^{\dagger}(x),i\phi_{r\pm F\sigma}(x)\right]=\\
-\sum_{q>0}\frac{1}{n_{q}}e^{-i\mathrm{sgn}(rF)q(x\mp x)}[b_{\sigma r\cdot q}^{\dagger},b_{\sigma r\cdot q}]=\\
\sum_{q>0}\frac{1}{n_{q}}e^{-i\mathrm{sgn}(rF)q(x\mp x)}.\end{multline*}
In total this leads to \begin{multline*}
A_{[r]_{b}[F]_{f^{-}}[\sigma]_{f^{+}}}(x):=e^{2\sum_{q>0}\frac{1}{n_{q}}\left(1-\cos(2qx)\right)}\\
\times e^{-2\sum_{q>0}\frac{1}{n_{q}}\left(\frac{\varepsilon_{0q}}{\varepsilon_{c+q}}+1\right)\sin^{2}(qx)}.\end{multline*}
Because of $\sin^{2}(qx)=\frac{1}{2}\left(1-\cos(2qx)\right)$ the
final result is \begin{equation}
A_{[r]_{b}[F]_{f^{-}}[\sigma]_{f^{+}}}(x):=e^{2\sum_{q>0}\frac{1}{n_{q}}\left(1-\frac{\varepsilon_{0q}}{\varepsilon_{c+q}}\right)\sin^{2}(qx)}.\end{equation}
 The same result is also obtained for all other processes with $S_{rF}=u$.

\section{\label{sec:Regularization-of-}Regularization of $\left\langle \vec{N}\vec{m}\left|V_{f^{+}\, b\, f^{-}}\right|\vec{N}\vec{m}\right\rangle $ }

As already mentioned in the main text, expression (\ref{eq:ME_V_nrr_expl})
for the matrix element $\left\langle \vec{N}\vec{m}\left|V_{S_{r}S_{F}S_{\sigma}}\right|\vec{N}\vec{'m}'\right\rangle $
diverges if $\sum_{j\delta q}\left|m_{j\delta q}-m'_{j\delta q}\right|\le1$
and if $V_{S_{r}S_{F}S_{\sigma}}$ is $\vec{N}$ conserving. Here
we show in detail how the matrix element can be properly regularized
for the case $\vec{m}=\vec{m}'$ and $V_{S_{r}S_{F}S_{\sigma}}=V_{f^{+}\, b\, f^{-}}$.
We start with equation (\ref{eq:ME_V_nrr_expl}),\begin{multline}
\left\langle \vec{N}\vec{m}\left|V_{f^{+}\, b\, f^{-}}\right|\vec{N}\vec{m}\right\rangle =\\
\frac{1}{4L}u^{+}\sum_{rF\sigma}\int dx\frac{e^{-2i\mathrm{sgn}(rF)(N_{r\sigma}-N_{r-\sigma})\frac{\pi}{L}x}}{4\sin^{2}\left(\frac{\pi}{L}x\right)}\\
\times\prod_{j\delta q}F(\tilde{\lambda}_{[r]_{f^{+}}[F]_{b}[\sigma]_{f^{-}}}^{j\delta q}(x),m_{j\delta q},m_{j\delta q}).\label{eq:VFSpBSFSmStart}\end{multline}
In a first step we rewrite the fraction $\frac{e^{-2i\mathrm{sgn}(rF)(N_{r\sigma}-N_{r-\sigma})\frac{\pi}{L}x}}{4\sin^{2}(\frac{\pi}{L}x)}$
as $\frac{e^{-2i\mathrm{sgn}(rF)N_{r\sigma}\frac{\pi}{L}x}}{1-e^{i\frac{2\pi}{L}x}}\frac{e^{2i\mathrm{sgn}(rF)N_{r-\sigma}\frac{\pi}{L}x}}{1-e^{-i\frac{2\pi}{L}x}}$
and, by using the identity \[
\sum_{n=-\infty}^{N}e^{-inx}=\frac{e^{-iNx}}{1-e^{ix}},\]
 we transform it into the product of two infinite sums extending over
the whole Fermi sea,\begin{multline}
\frac{e^{-2i\mathrm{sgn}(rF)(N_{r\sigma}-N_{r-\sigma})\frac{\pi}{L}x}}{4\sin^{2}(\frac{\pi}{L}x)}=\\
\sum_{n=-\infty}^{N_{r\sigma}}e^{-2i\mathrm{sgn}(rF)n\frac{\pi}{L}x}\sum_{n'=-\infty}^{N_{r-\sigma}}e^{2i\mathrm{sgn}(rF)n'\frac{\pi}{L}x}.\label{eq:prodofinfsums}\end{multline}
An important observation is, that the multiplication with $e^{-in_{q}x}-e^{in_{q}x}=e^{-in_{q}x}\left(1-e^{2in_{q}x}\right)$
recasts the infinite sum $\sum_{n=-\infty}^{N}e^{-2inx}$ into a finite
sum, \begin{multline}
e^{-in_{q}x}\left(1-e^{2in_{q}x}\right)\sum_{n=-\infty}^{N}e^{-2inx}=\\
e^{-in_{q}x}\sum_{n=N-n_{q}+1}^{N}e^{-2inx}.\label{eq:inftofin}\end{multline}
We now have a closer look at the coefficients $\tilde{\lambda}_{[r]_{f^{+}}[F]_{b}[\sigma]_{f^{-}}}^{j\delta q},$
which according to (\ref{eq:lambda_tilde}) are given by \begin{multline*}
\tilde{\lambda}_{[r]_{f^{+}}[F]_{b}[\sigma]_{f^{-}}}^{j\delta q}(x)=\\
\frac{1}{\sqrt{n_{q}}}\left[\Lambda_{r\sigma}^{j\delta}\left(e^{-i\mathrm{sgn}(rF)qx}-e^{i\mathrm{sgn}(rF)qx}\right)\right.\\
+\left.\Lambda_{r-\sigma}^{j\delta}\left(e^{i\mathrm{sgn}(rF)qx}-e^{-i\mathrm{sgn}(rF)qx}\right)\right].\end{multline*}
Then because of (\ref{eq:inftofin}), the product \begin{multline*}
\prod_{j\delta q}\left(\tilde{\lambda}_{[r]_{f^{+}}[F]_{b}[\sigma]_{f^{-}}}^{j\delta q}(x)\right)^{t_{j\delta q}}\\
\times\sum_{n=-\infty}^{N_{r\sigma}}e^{-2i\mathrm{sgn}(rF)n\frac{\pi}{L}x}\sum_{n'=-\infty}^{N_{r-\sigma}}e^{2i\mathrm{sgn}(rF)n'\frac{\pi}{L}x},\, r_{j\delta q}\in\mathbb{N}\end{multline*}
is a finite sum for $\sum_{j\delta q}t_{j\delta q}\ge2$. But from
(\ref{eq:Fvonlambda1}) we can conclude that \[
\prod_{j\delta q}F(\tilde{\lambda}_{[r]_{f^{+}}[F]_{b}[\sigma]_{f^{-}}}^{j\delta q}(x),m_{j\delta q},m_{j\delta q})=1+\mathcal{O}(\lambda^{2}),\]
where $\mathcal{O}(\lambda^{2})$ collects all those terms which contain
a factor $\prod_{j\delta q}\left(\tilde{\lambda}_{[r]_{f^{+}}[F]_{b}[\sigma]_{f^{-}}}^{j\delta q}(x)\right)^{t_{j\delta q}}$
with $\sum_{j\delta q}t_{j\delta q}\ge2.$ Thus \begin{multline}
\int dx\frac{e^{-2i\mathrm{sgn}(rF)(N_{r\sigma}-N_{r-\sigma})\frac{\pi}{L}x}}{4\sin^{2}\left(\frac{\pi}{L}x\right)}\\
\times\left(\prod_{j\delta q}F(\tilde{\lambda}_{[r]_{f^{+}}[F]_{b}[\sigma]_{f^{-}}}^{j\delta q}(x),m_{j\delta q},m_{j\delta q})-1\right)\label{eq:nondiverging_part}\end{multline}
 is a well defined integral over a finite sum and therefore not diverging.
On the other hand we find with (\ref{eq:prodofinfsums}) \begin{multline*}
\int dx\frac{e^{-2i\mathrm{sgn}(rF)(N_{r\sigma}-N_{r-\sigma})\frac{\pi}{L}x}}{4\sin^{2}\left(\frac{\pi}{L}x\right)}=\\
\int dx\sum_{n=-\infty}^{N_{r\sigma}}e^{-2i\mathrm{sgn}(rF)n\frac{\pi}{L}x}\sum_{n'=-\infty}^{N_{r-\sigma}}e^{2i\mathrm{sgn}(rF)n'\frac{\pi}{L}x}=\\
L\sum_{n=-\infty}^{N_{r\sigma}}\sum_{n'=-\infty}^{N_{r-\sigma}}\delta_{n,n'}=\sum_{n=-\infty}^{\min(N_{r\sigma},N_{r-\sigma})}L.\end{multline*}
Regularization of the previous expression now is easily achieved by
subtracting in the previous equation e.g. the contribution from below
half-filling, such that,\begin{equation}
\int dx\frac{e^{-2i\mathrm{sgn}(rF)(N_{r\sigma}-N_{r-\sigma})\frac{\pi}{L}x}}{4\sin^{2}\left(\frac{\pi}{L}x\right)}=L\min(N_{r\sigma},N_{r-\sigma}).\label{eq:regularized_part}\end{equation}
Combining (\ref{eq:nondiverging_part}) and (\ref{eq:regularized_part})
we obtain the finite expression,\begin{multline*}
\left\langle \vec{N}\vec{m}\left|V_{f^{+}\, b\, f^{-}}\right|\vec{N}\vec{m}\right\rangle =u^{+}\sum_{r}\min(N_{r\uparrow},N_{r\downarrow})\\
+\frac{1}{4L}u^{+}\sum_{rF\sigma}\int dx\frac{e^{-2i\mathrm{sgn}(rF)(N_{r\sigma}-N_{r-\sigma})\frac{\pi}{L}x}}{4\sin^{2}\left(\frac{\pi}{L}x\right)}\\
\times\left(\prod_{j\delta q}F(\tilde{\lambda}_{[r]_{f^{+}}[F]_{b}[\sigma]_{f^{-}}}^{j\delta q}(x),m_{j\delta q},m_{j\delta q})-1\right),\end{multline*}
which is equivalent to equation (\ref{eq:VBSii_mm}) in the main text.
The regularization for the case $\sum_{j\delta q}\left|m_{j\delta q}-m'_{j\delta q}\right|=1$
as well as for the matrix elements of the $\vec{N}$ conserving processes
$V_{f^{-}bf}$ and $V_{bf^{-}f^{+}}$ which are only relevant near
half-filling can be achieved in a similar way.

\end{document}